%% file: template.tex
\documentclass{article}

\usepackage{graphicx}
\usepackage{subfiles}
\usepackage{commath}
\usepackage{booktabs}
\usepackage{color}
\usepackage{amsfonts,amssymb,amsbsy,amsmath,amsthm}
\usepackage{bm}
\usepackage{authblk}

\renewcommand{\vec}[1]{\bm{#1}}
\newcommand{\mtrx}[1]{\vec{#1}}

\renewcommand{\d}{\partial}
\newcommand{\dd}{\, \mathrm{d}}

\DeclareMathOperator*{\LT}{LT}

\DeclareMathOperator*{\E}{E}
\DeclareMathOperator*{\Jac}{Jac}
\DeclareMathOperator*{\Hess}{Hess}

\theoremstyle{plain}
\newtheorem{theorem}{Theorem}[section]

\newtheorem{proposition}[theorem]{Proposition}

\theoremstyle{definition}
\newtheorem{definition}[theorem]{Definition}
\newtheorem{example}[theorem]{Example}

\theoremstyle{remark}
\newtheorem*{remark}{Remark}

\begin{document}

\title{Holonomic extended least angle regression}

\author{Marc H\"ark\"onen \thanks{School of Mathematics, Georgia Institute of Technology,}\and Tomonari Sei \thanks{Graduate School of Information Science and Technology, The University of Tokyo} \and Yoshihiro Hirose \thanks{Graduate School of Information Science and Technology, Hokkaido University}
}
\maketitle

\begin{abstract}
  One of the main problems studied in statistics is the fitting of models.
  Ideally, we would like to explain a large dataset with as few parameters as possible.
  There have been numerous attempts at automatizing this process. Most notably, the Least Angle Regression algorithm, or LARS, is a computationally efficient algorithm that ranks the covariates of a linear model.
The algorithm is further extended to a class of distributions in the generalized linear model by using properties of the manifold of exponential families as dually flat manifolds.
However this extension assumes that the normalizing constant of the joint distribution of observations is easy to compute. This is often not the case, for example the normalizing constant may contain a complicated integral.
We circumvent this issue if the normalizing constant satisfies a holonomic system,
a system of linear partial differential equations with a finite-dimensional space of solutions.
In this paper we present a modification of the holonomic gradient method and add it to the extended LARS algorithm. We call this the holonomic extended least angle regression algorithm, or HELARS. The algorithm was implemented using the statistical software \texttt{R}, and was tested with real and simulated datasets.
\end{abstract}

\subfile{Sections/intro.tex}

\subfile{Sections/statistics.tex}

\subfile{Sections/holonomic.tex}

\subfile{Sections/bisect_regr.tex}

\subfile{Sections/truncated_normal.tex}

\subfile{Sections/discussion.tex}

\bibliographystyle{spmpsci}      
\bibliography{references.bib}

\end{document}


\section{Submanifolds of the exponential family}
\label{app:submanif}
\begin{definition}
	Let $S$ and $M$ be manifolds, with $M\subseteq S$, and coordinate systems $\vec\xi = (\xi^1,\dotsc,\xi^n)$ and $\vec\theta = (\theta^1, \dotsc, \theta^d)$ respectively. $M$ is a \emph{submanifold} of $S$ is
	\begin{enumerate}
		\item The restriction $\left.\xi_i\right|_M$ of each $\xi_i \colon S \to \mathbb{R}$ is a $C^\infty$ function on $M$ for each $1\leq i \leq n$.

		\item Let $B_a^i := \left( \frac{\partial \xi^i}{\partial \theta^a} \right)_p$, and $B_a := (B_a^1,\dotsc, B_a^n) \in \mathbb{R}^n$ for all $1\leq a \leq d$ and $1\ leq i \leq n$. Then for each point $p\in M$, $B_1,\dotsc,B_d$ are linearly independent.

		\item For any open subset $W$ of $M$, there exists $U$, an open subset of $S$, such that $W = M \cap U$.
	\end{enumerate} \label{def:submanif}
\end{definition}

As in \Cref{sub:dually_flat_manifolds}, let $S$ be the manifold of the exponential family
\begin{align*}
	S=\{ p(\vec y \mid \vec\xi) = \exp(C(\vec y) + \vec\xi \cdot \vec F(\vec y) - \psi(\vec\xi)) \mid \xi \in \mathbb{R}^n\}
\end{align*}
Now we prove that the subsets appearing in \Cref{prop:submanifold_fix_parameters,prop:submanifold_linear_transform} are indeed submanifolds by checking each condition in \Cref{def:submanif}.
\begin{proof}[Proof of \Cref{prop:submanifold_fix_parameters}.]
Let $I \subset \{1,2,\dotsc,n\}$ be a nonempty set and let $\{c_i\}_{i\in I}$ be a set of fixed real numbers indexed by $I$. Let
	\begin{align*}
		M = \{\vec\xi \mid \xi^i = c_i\ \forall i\in I\}.
	\end{align*}
	Now $(\xi^j)_{j\not\in I}$ is a coordinate system for $M$.
\begin{enumerate}
	\item For all $i \in I$, the restriction $\left.\xi_i\right|_M = c_i$ is a constant, which is $C^\infty$. If $i\not\in I$, then $\left.\xi^i\right|_M = \xi^i$, which is $C^\infty$ by definition of a coordinate system.

	\item Note that $B_a^i = \delta_a^i$, so $B_a$ is a vector in $\mathbb{R}^n$, where the $a$th element is 1 and all others are 0. Clearly $B_1,\dotsc,B_d$ are linearly independent.

	\item Let $W$ open in $M$ and let $\varepsilon > 0$. Define the set $U$ as
	\begin{align*}
		U = \{\vec\xi \mid \xi^j \in W ~ \forall j\not\in I, ~~\xi^i \in (c_i - \varepsilon, c_i + \varepsilon) ~\forall i\in I\}.
	\end{align*}
	Then $U$ is open and $W = U \cap M$.
\end{enumerate}
\end{proof}

\begin{proof}[Proof of \Cref{prop:submanifold_linear_transform}.]
Let $d < n$, and let $\mtrx X$ be an $(n \times d)$ matrix of rank $d$, and
	\begin{align*}
		M = \{ \vec\xi \mid \vec\xi = \mtrx X \vec\theta,~\vec\theta \in \mathbb{R}^d \}.
	\end{align*}
	The vector $\vec\theta$ is a coordinate system for $M$.
	\begin{enumerate}
		\item The restriction is
		\begin{align*}
			\left.\xi_i\right|_M = \sum_{a=1}^d X_{ia} \theta^a,
		\end{align*}
		which is clearly $C^\infty$.

		\item Here $B_a^i = \frac{\partial\xi^i}{\partial\theta^a} = X_{ia}$, so the vector $B_a = \vec X_a$, the $a$th column of $\mtrx X$. Since we assumed that $\mtrx X$ is of full rank, all columns are linearly independent.

		\item Let $W$ be open in $M$. Define $U = (S\setminus M) \cup W$. Then any point on $S\setminus M$ has an open neighbourhood in $S$ since $M$ is closed in $S$. Any point on $W$ has an open $d$ dimensional ball in $M$. Let $\delta$ be the diameter of the image of this ball on $S$. For a small $\varepsilon > 0$, there is a $n$ dimensional ball of diameter $\delta - \varepsilon$ fully contained in $(S\setminus M) \cup W$. Hence $U$ is open and $U\cap M = W \cap M = W$.
	\end{enumerate}
\end{proof}
\clearpage

\section{Code, (non-holonomic) bisector regression}
Sample code for the diabetes data in \cite{hirose2010extension}. Files for the source code and dataset can be found in \url{http://people.math.gatech.edu/~mharkonen3/helars.html}.
\label{app:code_elars}

\lstset{
language=R,         
basicstyle=\tiny,          
columns=fullflexible
}
\begin{lstlisting}
library(Matrix)
library(nleqslv)

Diabetes <- read.table("diabetes.data", header = T)

# Scale
Diabetes[-11] <- scale(Diabetes[-11])[ , ]
Diabetes[11] <- scale(Diabetes[11], scale = FALSE)[ , ]
attach(Diabetes)

# Useful global variables
n <- nrow(Diabetes) # Sample size
p <- ncol(Diabetes) - 1 # Parameters
r <- 1 # Depends on the model

# Design matrix
X <- as.matrix(Diabetes[-11])
X.tilde <- cbind(1,X)
X.B <- as.matrix(bdiag(X.tilde, diag(rep(1,r))))

# Response
y <- Diabetes$Y

# Convert theta to xi
theta2xi <- function(theta) {
 if (length(theta) != p+r+1) {
	print(paste("theta2xi error, argument needs size ", p+r+1))
	return(0)
 }

 return(X.B %*% theta)
}

# Convert xi to mu
# MODIFY FOR EACH MODEL
xi2mu <- function(xi) {
 if (length(xi) != n+r) {
	print(paste("xi2mu error, argument needs size ", n+r))
	return(0)
 }
 mu <- -xi[1:442]/(2*xi[443])
 mu <- c(mu, sum(xi[1:442]^2)/(4*xi[443]^2) - 442/(2*xi[443]))
 return(mu)
}

# Convert mu to xi
# MODIFY FOR EACH MODEL
mu2xi <- function(mu) {
 if (length(mu) != n+r) {
	print(paste("mu2xi error, argument needs size ", n+r))
	return(0)
 }

 xi <- 442*mu[1:442]/(mu[443] - sum(mu[1:442]^2))
 xi <- c(xi, -442/(2*(mu[443] - sum(mu[1:442]^2))))
 return(xi)
}

# Convert from mu to eta
mu2eta <- function(mu) {
 if (length(mu) != n+r) {
	print(paste("mu2eta error, argument needs size ", n+r))
	return(0)
 }

 return(t(X.B)%*%mu)
}

theta2eta <- function(theta) {
 if (length(theta) != p+r+1) {
	print(paste("theta2eta error, argument needs size ", p+r+1))
	return(0)
 }

 return (mu2eta(xi2mu(theta2xi(theta))))
}

eta2theta <- function(eta) {
 if (length(eta) != p+r+1) {
	print(paste("eta2theta error, argument needs size ", p+r+1))
	return(0)
 }

 Y <- ginv(X.B) # Y is the generalized inverse of X.B

 return (Y %*% mu2xi(t(Y) %*% eta))
}

# MODIFY FOR EACH MODEL
psi.star <- function(xi) {
 if (length(xi) != n+r) {
	print(paste("psi.star error, argument needs size ", p+r+1))
	return(0)
 }

 xi <- -sum(xi[1:442]^2)/(4*xi[443]) - 442/2*log(-xi[443]) + 442/2*log(pi)
 return(xi)
}

phi.star <- function(mu) {
 if (length(mu) != n+r) {
	print(paste("phi.star error, argument needs size ", n+r))
	return(0)
 }

 return (sum(mu2xi(mu)*mu) - psi.star(mu2xi(mu)))
}

psi <- function(theta) {
 if (length(theta) != p+r+1) {
	print(paste("psi error, argument needs size ", p+r+1))
	return(0)
 }

 return (psi.star(theta2xi(theta)))
}

psi.I <- function(theta,I) {
 if (length(theta) != p+r+1 || length(I) != p+r+1) {
	print(paste("psi.I error, argument needs size ", p+r+1))
	return(0)
 }

 return (psi(theta*I))
}

phi <- function(theta) {
 if (length(theta) != p+r+1) {
	print(paste("phi error, argument needs size ", p+r+1))
	return(0)
 }

 return (sum(theta2eta(theta) * theta) - psi(theta))
}

phi.I <- function(theta, I) {
 if (length(theta) != p+r+1 || length(I) != p+r+1) {
	print(paste("phi.I error, argument needs size ", p+r+1))
	return(0)
 }

 return (sum(theta2eta(theta) * theta * I) - psi.I(theta, I))
}

# Divergence in M(I)
div.I <- function(theta1, theta2, I=!logical(p+r+1)) {
 return(phi.I(theta1, I) + psi.I(theta2, I) - sum(theta2eta(theta1) * theta2 * I))
}

# Fisher information matrix, i.e. second derivatives of psi.star
# MODIFY FOR EACH MODEL
fisher <- function(xi) {
 if (length(xi) != n+r) {
	print(paste("fisher error, argument needs size ", n+r))
	return(0)
 }

 inf <- matrix(nrow = 443, ncol = 443)
 for (i in 1:442) {
	for (j in 1:442) {
	 inf[i,j] = 0
	}
	inf[i,i] = -1/(2*xi[443])
	inf[i,443] = xi[i]/(2*xi[443]^2)
	inf[443,i] = inf[i,443]
 }
 inf[443,443] = -sum(xi[1:442]^2)/(2*xi[443]^3) + 442/(2*xi[443]^2)
 return(inf)
}

# Compute the Jacobian of the transformation eta to theta
jacob <- function(theta) {
 return(t(X.B) %*% fisher(theta2xi(theta)) %*% X.B)
}

# Convert mixed coordinates to theta. idx.eta is a logical vector, where TRUE position
# corresponds to a eta-coordinate in mix, and false corresponds to 
# an theta-coordinate
mixed2theta <- function(mix, idx.eta, theta.guess = theta1) {
 # Function whose root we want to find. The argument vars refers to the
 # unknown thetas in mix (i.e. theta[idx.eta])
 froot <- function(vars) {
	new.theta <- mix
	new.theta[idx.eta] <- vars
	return(mix[idx.eta] - theta2eta(new.theta)[idx.eta])
 }

 # Jacobian of froot. Again, vars contain the unknown thetas
 fjac <- function(vars) {
	new.theta <- mix
	new.theta[idx.eta] <- vars
	G <- jacob(new.theta)
	# Return a submatrix of G. The element (i,j) belongs to the submatrix
	# if both idx.eta[i]==TRUE and idx.eta[j]==TRUE
	return(-G[which(idx.eta), which(idx.eta)])
 }
 # First guess at the MLE of zero model
 guess <- theta.guess[idx.eta]

 # Numerical solver may need tweaking
 res <- nleqslv(guess,froot, fjac, xscalm="auto", global = "none", control=list(xtol=1e-16))
 print(res$message)
 if(res$termcd != 1) browser()
 mix[idx.eta] <- res$x
 return(mix)
}

# Compute theta coordinates of the m-projection of theta to M(i,a,I)
m.projection <- function(theta, i, I, a=0) {
 eta <- theta2eta(theta)
 mix <- eta
 mix[!I] = 0
 mix[i] = a
 I[i] = FALSE
 return(mixed2theta(mix,I, theta))
}

# Get the value of component i such that the divergence from cur.theta is
# equal to t.star
ger.component <- function(t.star, i, I, cur.theta = theta1) { 
 r <- c(0, cur.theta[i])
 for(k in 1:10) {
	tmp = mean(r)
	theta.hat <- m.projection(cur.theta, i, I, tmp)
	if (div.I(cur.theta, theta.hat, I) - t.star < 0) {
	 r[2] <- tmp
	} else {
	 r[1] <- tmp
	}
 }
 return (mean(r))
}

# MLE of the full model
fit <- glm(y ~ X, family = gaussian(link = "identity"))
s2 <- sum(fit$residuals^2)/fit$df.residual # Estimate of the variance
theta1 <- c(fit$coefficients/s2, -1/(2*s2))
names(theta1)[12] <- "S2"

# MLE of the zero model
fit0 <- glm(y ~ 1, family = gaussian(link = "identity"))
s20 <- sum(fit0$residuals^2)/fit0$df.residual # Estimate of the variance
theta0 <- c(coefficients(fit0)/s20, rep(0,10), -1/(2*s20))
names(theta0) <- names(theta1)
eta0 <- theta2eta(theta0)

main <- function() {
 theta <- theta1
 df <- data.frame(t(theta))
 maxdiv <- div.I(theta,theta0)
 df <- cbind(df, 1)
 names(df)[length(theta)+1] <- "Div/max(Div)"

 # Main loop
 I <- !logical(p+r+1) # Variables present
 for (k in 1:p) {
	I.small <- I[(1:p)+1] # Active theta 1 to d
	nn <- sum(I.small) # Number of active covariates
	t <- rep(0,nn)
	ind <- 1
	for (i in which(I.small)+1) { # For every active covariate, compute m-proj and divergence
	 theta.bar <- m.projection(theta, i, I)
	 t[ind] <- div.I(theta,theta.bar,I)
	 ind <- ind+1

	 print(paste(k,i))
	}
	t.star <- min(t)
	i.star <- which(I.small)[which.min(t)]

	I.small[i.star] <- FALSE

	print("Get rest of components")
	theta.next <- theta
	for (i in which(I.small)+1) {
	 theta.next[i] <- ger.component(t.star, i, I, theta)
	}
	theta.next[i.star + 1] <- 0

	print("Wrapup step")
	# Find theta[0] and theta[12] given eta[0], eta[12]
	mix<-c(eta0[1], theta.next[(1:p)+1])
	if (r>0) mix <- c(mix, eta0[(1:r)+p+1])
	II <- logical(p+r+1)
	II[1] <- TRUE
	if(r > 0) II[(1:r)+p+1] <- TRUE
	theta <- mixed2theta(mix,II,theta)

	#theta <- theta.next
	I[i.star+1] <- FALSE
	df <- rbind(c(theta, div.I(theta,theta0)/maxdiv),df)
 }
 
 return(df)
}
 ### PLOTTING
 n.thetas <- p+r+1
 plot(df[,n.thetas+1], df[,2], ylim=c(min(df[,2:n.thetas]), max(df[,2:n.thetas])),
	xlab = "div/max(div)", ylab="value of parameter", main="ELARS, Diabetes data")
 lines(df[,n.thetas+1], df[,2])
 for (k in (2:p)+1) {
	lines(df[,n.thetas+1], df[,k], type="o")
 }
 abline(v=df[ ,n.thetas+1])
 text(1.02, df[p+1, 1:p + 1], 1:p)
 
 ############

get.order <- function(df) {
 dims <- dim(df)
 I <- !logical(dims[2])
 I[c(1, dims[2])] <- FALSE
 order <- NULL
 for(i in (dims[1]-1):1) {
	zero <- which(df[i,] == 0 & I)
	order <- c(order, zero)
	I[zero] <- FALSE
 }
 try(print(colnames(X)[order]))
 return(order - 1)
}
\end{lstlisting}

\clearpage
\section{Holonomic ELARS implementation for the truncated normal distribution} \label{app:helars_code}
Implementation of the Holonomic ELARS algorithm, using the diabetes dataset in \cite{hirose2010extension}. Files for the source code and dataset can be found in \url{http://people.math.gatech.edu/~mharkonen3/helars.html}.

\begin{lstlisting}
library("Matrix")
library("hgm")
library("deSolve")
library("nleqslv")

Diabetes <- read.table("diabetes.data", header = T)

X.unnorm <- Diabetes[,-11]
X.norm <- scale(X.unnorm)
X <- cbind(1,X.norm[,])
centers <- attr(X.norm, "scaled:center")
scales <- attr(X.norm, "scaled:scale")

X.B <- as.matrix(bdiag(X,1))

sample <- Diabetes[,11]

Y <- c(sample, sum(sample^2))

# Useful global variables
n <- length(Y) - 1 # Sample size
p <- ncol(X) - 1 # Parameters
r <- 1 # Depends on the model

##### MLE ####
## MLE of the full model
mle.full <- function(X.B, Y) {
 theta0 <- c(rep(0, p+1) ,-1) #Initial guess
 l <- length(theta0)
 log.A <- rep(log(sqrt(pi)/2), n) #Initial log.A
 theta <- theta0 
 gamma <- 0.1 # Gradient descent step size

 for(i in 1:1000) {
	grad <- t(Y - grad.psi(X.B%*%theta, log.A)) %*% X.B # Gradient of log-likelihood
	grad <- t(grad)
	hess <- hess.psi(X.B%*%theta, log.A) 
	hess <- -t(X.B)%*%hess%*%X.B # Hessian of log-likelihood

	diff <- solve(hess, -grad) # Newton's method

	if(diff[l]+theta[l] > 0) { # If Newton's method fails, use gradient descent
	 print("Grad desc")
	 diff <- gamma*grad/sqrt(crossprod(grad))[1]
	}

	theta1 <- theta + 0.1*diff # Slow down Newton's method for more accurate log.A
	log.A <- update.log.A(theta, log.A, theta1)
	theta <- theta1
	print(paste("Iteration number", i))
	print(loglike(theta, log.A))
	print(theta)
	if(crossprod(grad) < 1e-12) break

 }
 return(list(theta = theta, log.A = log.A))
 ## Start with grad descent, switch to newton after a few iteraitons
}

## Mle of the simplest model (theta_a = 0)
mle.simplest <- function(X.B, Y) {
 theta0 <- c(rep(0, p+1),-1) # Initial guess
 l <- length(theta0)
 log.A <- rep(log(sqrt(pi)/2), n) # Initial log.A
 theta <- theta0
 gamma <- 0.1 # Gradient descent step size

 for(i in 1:1000) {
	grad <- t(Y-grad.psi(X.B%*%theta, log.A)) %*% X.B # Gradient of log-likelihood
	grad <- t(grad)
	hess <- hess.psi(X.B%*%theta, log.A)
	hess <- -t(X.B)%*%hess%*%X.B # Hessian of log-likelihood
	
	grad <- grad[c(1,l)]
	hess <- hess[c(1,l), c(1,l)]

	diff <- solve(hess, -grad)

	if(diff[2]+theta[l] > 0) { # If Newton's method fails, use gradient descent
	 print("Grad desc")
	 diff <- gamma*grad/sqrt(crossprod(grad))[1]
	}

	diff <- c(diff[1],rep(0,p), diff[2])

	theta1 <- theta+0.1*diff # Slow down Newton's method for more accurate log.A
	log.A <- update.log.A(theta, log.A, theta1)
	theta <- theta1
	print(paste("Iteration number", i))
	print(loglike(theta, log.A))
	print(theta)
	if(crossprod(grad) < 1e-12) break

 }
 return(list(theta = theta, log.A = log.A))
}

##### USEFUL FUNCTIONS FOR HOLONOMIC

# Use HGM to get new log.A given old theta and old log.A
update.log.A <- function(theta.old, log.A, theta.new, iter=1) {
 dF <- function(theta, log.A) {
	return(t(grad.log.A.theta(theta, log.A)))
 }
 new <- hgm.Rhgm(as.vector(theta.old), log.A, as.vector(theta.new), dF, 0:iter/iter)
 dd <- dim(new)
 return(as.vector(new[dd[1], -1]))
 return(new)
}

# Use HGM to get new log.A given old xi and old log.A
update.log.A.xi <- function(xi.old, log.A, xi.new) {
 new.log.A <- log.A
 for (i in 1:n) {
	new.log.A[i] <- hgm.Rhgm(xi.old[c(i,n+1)], log.A[i], xi.new[c(i, n+1)], grad.log.A)
 }
 return(new.log.A)
}

# Use HGM to get new log.A given old mixed coordinates and old log.A
update.log.A.mixed <- function(mix.old, log.A, mix.new, idx.eta, iter = 1, mix.guess = NULL) {
 I <- idx.eta
 theta.old <- mixed2theta(mix.old, idx.eta, log.A, mix.guess)$theta
 eta.old <- theta2eta(theta.old, log.A)
 mix.guess <- theta.old
 mix.guess[!idx.eta] <- eta.old[!idx.eta]
 dF <- function(mix, log.A) {
	theta <- mixed2theta(mix, I, log.A, mix.guess)$theta
	xi <- theta2xi(theta)
	l <- length(mix)
	n <- length(log.A)

	dldt <- grad.log.A.theta(theta, log.A)
	dmdx <- fisher(xi, log.A)
	didt <- t(X.B) %*% dmdx %*% X.B
	dtdi <- solve(didt)

	b <- matrix(0, nrow = sum(I), ncol = l)
	b[,I] <- diag(sum(I))
	b[,!I] <- -didt[I,!I]

	dm2tdmix.sub <- solve(didt[I,I], b)
	dm2tdmix <- matrix(0, nrow = l, ncol = l)
	dm2tdmix[I,] <- dm2tdmix.sub
	dm2tdmix[!I,!I] <- diag(sum(!I))
	return(t(dldt %*% dm2tdmix))
	# return(dm2tdmix)
 }

 new <- hgm.Rhgm(as.vector(mix.old), log.A, as.vector(mix.new), dF, 0:iter/iter)
 dd <- dim(new)
 return(as.vector(new[dd[1], -1]))
}

# Gradient of A_a
grad.log.A <- function(xi, log.A) {
 if(length(xi) != 2) stop("grad.log.A error, check argument length")
 r <- rep(0,2)
 r[1] <- -1/(2*xi[2])*(exp(-log.A)+xi[1])
 r[2] <- -1/(2*xi[2])*(1+xi[1]*r[1])
 return(r)
}

# Gradient of log.A wrt theta
grad.log.A.theta <- function(theta, log.A) {
	l <- length(log.A)
	d <- matrix(0, nrow = l, ncol = l+1)
	xi <- theta2xi(theta)
	for(i in 1:l) {
	 d[i, c(i,l+1)] <- grad.log.A(xi[c(i, l+1)], log.A[i])
	}
	return(d %*% X.B)
}

# Gradient of the potential function wrt xi
grad.psi <- function(xi, log.A) {
 r <- rep(0,n+1)
 for(i in 1:n) {
	glog <- grad.log.A(xi[c(i,n+1)], log.A[i])
	r[i] <- glog[1]
	r[n+1] <- r[n+1] + glog[2]
 }
 return(r)
}

# Hessian of the potential function wrt xi
hess.psi <- function(xi, log.A) {
 h <- matrix(0, nrow = n+1, ncol = n+1)
 for(i in 1:n) {
	xi.cur <- xi[c(i,n+1)]
	d <- rep(0,4)
	d[c(1,2)] <- grad.log.A(xi.cur, log.A[i])
	d[3] <- -1/(2*xi.cur[2])*(2*d[1] + xi.cur[1]*d[2])
	d[4] <- -1/(2*xi.cur[2])*(3*d[2] + xi.cur[1]*d[3])
	h[i,i] <- d[2] - d[1]^2
	h[i,n+1] <- d[3] - d[1]*d[2]
	h[n+1,i] <- h[i,n+1]
	h[n+1,n+1] <- h[n+1,n+1] + d[4] - d[2]^2
 }
 return(h)
}

# Log-likelihood
loglike <- function(theta, log.A) {
 return(t(Y)%*%X.B%*%theta - sum(log.A))
}

# DEBUG/SLOW: Compute log.A given theta
get.log.A <- function(theta, nsteps = 100) {
 theta0 <- c(rep(0,p+1), -1)
 log.A <- rep(log(sqrt(pi)/2), n)
 theta.old <- theta0
 for(i in 1:nsteps) {
	theta.next <- theta0 + i/nsteps*(theta - theta0)
	log.A <- update.log.A(theta.old, log.A, theta.next)
	theta.old <- theta.next
 }
 return(log.A)
}

# DEBUG/SLOW: Compute log.A given xi
get.log.A.xi <- function(xi, nsteps = 100) {
 xi0 <- c(rep(0,n), -1)
 log.A <- rep(log(sqrt(pi)/2), n)
 xi.old <- xi0
 for(i in 1:nsteps) {
	xi.next <- xi0 + i/nsteps*(xi - xi0)
	log.A <- update.log.A.xi(xi.old, log.A, xi.next)
	xi.old <- xi.next
 }
 return(log.A)
}

# Jacobian of log.A wrt xi
J.log.A <- function(xi, log.A) {
 J <- matrix(0, ncol = n+1, nrow = n)
 for (i in 1:n) {
	J[i, c(i,n+1)] <- grad.log.A(xi[c(i, n+1)], log.A[i])
 }
 return(J)
}

##### END USEFUL FUNCTIONS FOR HOLONOMIC

# Convert theta to xi
theta2xi <- function(theta) {
 if (length(theta) != p+r+1) {
	print(paste("theta2xi error, argument needs size ", p+r+1))
	return(0)
 }

 return(X.B %*% theta)
}

# Convert xi to mu
# MODIFY FOR EACH MODEL
xi2mu <- function(xi, log.A) {
 if (length(xi) != n+r) {
	print(paste("xi2mu error, argument needs size ", n+r))
	return(0)
 }
 
 mu <- grad.psi(xi, log.A)
 return(mu)
}

# Convert mu to xi
# MODIFY FOR EACH MODEL
mu2xi <- function(mu, log.A) {
 if (length(mu) != n+r) {
	print(paste("mu2xi error, argument needs size ", n+r))
	return(0)
 }
 xi <- mu
 xi[n+1] <- 1/(2*sum(mu[-(n+1)]^2) - 2*mu[n+1])*(n - sum(mu[-(n+1)] / exp(log.A)))
 xi[-(n+1)] <- -1/exp(log.A) - 2*xi[n+1]*mu[-(n+1)]

 return(xi)

}

# Convert from mu to eta
mu2eta <- function(mu) {
 if (length(mu) != n+r) {
	print(paste("mu2eta error, argument needs size ", n+r))
	return(0)
 }

 return(t(X.B)%*%mu)
}

theta2eta <- function(theta, log.A) {
 if (length(theta) != p+r+1) {
	print(paste("theta2eta error, argument needs size ", p+r+1))
	return(0)
 }

 return (mu2eta(xi2mu(theta2xi(theta), log.A)))
}

eta2theta <- function(eta, log.A) {
 if (length(eta) != p+r+1) {
	print(paste("eta2theta error, argument needs size ", p+r+1))
	return(0)
 }

 Y <- ginv(X.B) # Y is the generalized inverse of X.B
 foo <- mu2xi(t(Y) %*% eta, log.A)
 xi <- foo

 return (Y %*% xi)
}

# MODIFY FOR EACH MODEL
psi.star <- function(xi, log.A) {
 if (length(xi) != n+r) {
	print(paste("psi.star error, argument needs size ", p+r+1))
	return(0)
 }

 return(sum(log.A))
}

phi.star <- function(mu, log.A) {
 if (length(mu) != n+r) {
	print(paste("phi.star error, argument needs size ", n+r))
	return(0)
 }

 xi <- mu2xi(mu, log.A)
 return (sum(xi*mu) - psi.star(xi, log.A))
}

psi <- function(theta, log.A) {
 if (length(theta) != p+r+1) {
	print(paste("psi error, argument needs size ", p+r+1))
	return(0)
 }

 return (psi.star(theta2xi(theta), log.A))
}

psi.I <- function(theta, I, log.A) {
 if (length(theta) != p+r+1 || length(I) != p+r+1) {
	print(paste("psi.I error, argument needs size ", p+r+1))
	return(0)
 }

 log.A <- update.log.A(theta, log.A, theta*I)
 return (psi(theta*I, log.A))
}

phi <- function(theta, log.A) {
 if (length(theta) != p+r+1) {
	print(paste("phi error, argument needs size ", p+r+1))
	return(0)
 }

 return (sum(theta2eta(theta, log.A) * theta) - psi(theta, log.A))
}

phi.I <- function(theta, I, log.A) {
 if (length(theta) != p+r+1 || length(I) != p+r+1) {
	print(paste("phi.I error, argument needs size ", p+r+1))
	return(0)
 }

 log.A <- update.log.A(theta, log.A, theta*I)
 return (sum(theta2eta(theta, log.A) * theta * I) - psi.I(theta, I, log.A))
}

# Divergence in M(I)
div.I <- function(theta1, log.A1, theta2, log.A2, I=!logical(p+r+1)) {
 return(phi.I(theta1, I, log.A1) + psi.I(theta2, I, log.A2) - sum(theta2eta(theta1, log.A1) * theta2 * I))
}

# Fisher information matrix, i.e. second derivatives of psi.star
# MODIFY FOR EACH MODEL
fisher <- hess.psi

# Compute the Jacobian of the transformation eta to theta
jacob <- function(theta, log.A) {
 return(t(X.B) %*% fisher(theta2xi(theta), log.A) %*% X.B)
}

# Convert mixed coordinates to theta. idx.eta is a logical vector, where TRUE position
# corresponds to a eta-coordinate in mix, and false corresponds to 
# an theta-coordinate
mixed2theta <- function(mix, idx.eta, log.A, res.guess = NULL) {
 if(is.null(res.guess)) {
	theta <- c(rep(0,p+1), -1)
	eta <- theta2eta(theta,log.A)
 } else {
	eta <- mix
	eta[!idx.eta] <- res.guess[!idx.eta]
	theta <- mix
	theta[idx.eta] <- res.guess[idx.eta]
 }
 F <- function(res) {
	eta <- mix
	eta[!idx.eta] <- res[!idx.eta]
	theta <- mix
	theta[idx.eta] <- res[idx.eta]
	return(eta - theta2eta(theta, log.A))
 }

 H <- function(res) {
	eta <- mix
	eta[!idx.eta] <- res[!idx.eta]
	theta <- mix
	theta[idx.eta] <- res[idx.eta]

	hess <- t(X.B) %*% hess.psi(theta2xi(theta), log.A) %*% X.B

	result <- diag(p+r+1)
	result[,idx.eta] <- -hess[,idx.eta]
	return(result)
 }

 res <- nleqslv(fn = F, jacobian = H, x = res.guess)
 
 if(res$termcd == 1 || res$termcd == 2) {
	res <- res$x
 } else {
	res <- theta
	res[!idx.eta] <- eta[!idx.eta]

	l <- length(res)

	for(i in 1:1000) {
	 diff <- solve(H(res), -F(res))
	 res1 <- res + diff
	 if(idx.eta[l] == TRUE && res1[l] > 0) res1 <- res + 0.5^ceiling(log(-res[l]/diff[l])/log(0.5))*diff
	 res <- res1
	 if(sum(F(res)^2) < 1e-12) break
	}
 }

 eta <- mix
 eta[!idx.eta] <- res[!idx.eta]
 theta <- mix
 theta[idx.eta] <- res[idx.eta]
 return(list(res=res, theta=theta, eta=eta))
}

# Compute theta coordinates of the m-projection of theta to M(i,a,I)
m.projection <- function(theta, log.A, i, I, a=0, iter = 1) {
 eta <- theta2eta(theta, log.A)
 mix <- eta
 mix[!I] = 0
 mix[i] = a
 I[i] = FALSE
 mix.old <- mix
 mix.old[i] <- theta[i]

 mix.guess <- eta
 mix.guess[I] <- theta[I]

 log.A.new <- update.log.A.mixed(mix.old, log.A, mix, I, iter, mix.guess)

 res.guess <- eta
 res.guess[I] <- theta[I]

 theta.new <- mixed2theta(mix, I, log.A.new, res.guess)$theta
 return(list(theta = theta.new, log.A = log.A.new))
}

# Get the value of component i such that the divergence from cur.theta is
# equal to t.star
get.component <- function(t.star, i, I, cur.theta, log.A) { 
 r <- c(0, cur.theta[i])
 for(k in 1:3) {
	tmp <- mean(r)
	proj <- m.projection(cur.theta, log.A, i, I, tmp)
	theta.hat <- proj$theta
	log.A.hat <- proj$log.A
	if (div.I(cur.theta, log.A, theta.hat, log.A.hat, I) - t.star < 0) {
	 r[2] <- tmp
	} else {
	 r[1] <- tmp
	}
 }
 return (mean(r))
}

main <- function(MLE1, MLE0) {
 full <- MLE1
 theta <- full$theta
 log.A <- full$log.A
 simple <- MLE0
 theta0 <- simple$theta
 log.A0 <- simple$log.A
 eta0 <- theta2eta(theta0, log.A0)
 df <- data.frame(t(theta))
 maxdiv <- div.I(theta, log.A, theta0, log.A0)
 df <- cbind(df, 1)
 names(df)[length(theta)+1] <- "Div/max(Div)"

 # Main loop
 I <- !logical(p+r+1) # Variables present
 for (k in 1:p) {
	I.small <- I[(1:p)+1] # Active theta 1 to d
	nn <- sum(I.small) # Number of active covariates
	t <- rep(0,nn)
	ind <- 1
	for (i in which(I.small)+1) { # For every active covariate, compute m-proj and divergence
	 proj <- m.projection(theta, log.A, i, I, iter = 2)
	 theta.bar <- proj$theta
	 log.A.bar <- proj$log.A
	 t[ind] <- div.I(theta, log.A ,theta.bar, log.A.bar, I)
	 ind <- ind+1

	 print(paste(k,i))
	}
	t.star <- min(t)
	i.star <- which(I.small)[which.min(t)]

	I.small[i.star] <- FALSE

	print("Get rest of components")
	theta.next <- theta
	for (i in which(I.small)+1) {
	 theta.next[i] <- get.component(t.star, i, I, theta, log.A)
	 print(paste(i, "done"))
	}
	theta.next[i.star + 1] <- 0

	print("Wrapup step")
	# Find theta[0] and theta[12] given eta[0], eta[12]
	mix<-c(eta0[1], theta.next[(1:p)+1])
	if (r>0) mix <- c(mix, eta0[(1:r)+p+1])
	II <- logical(p+r+1)
	II[1] <- TRUE
	if(r > 0) II[(1:r)+p+1] <- TRUE

	mix.old <- theta
	mix.old[II] <- theta2eta(theta, log.A)[II]

	mix.guess <- theta
	mix.guess[!II] <- theta2eta(theta, log.A)[!II]

	log.A.next <- update.log.A.mixed(mix.old, log.A, mix, II,2, mix.guess)

	mix.res <- mixed2theta(mix,II, log.A.next, mix.guess)
	theta <- mix.res$theta
	log.A <- log.A.next

	I[i.star+1] <- FALSE
	df <- rbind(c(theta, div.I(theta, log.A, theta0, log.A0)/maxdiv),df)
 }
 
 return(df)
}
### PLOTTING
make.plot <- function(df) {
 n.thetas <- p+r+1
 plot(df[,n.thetas+1], df[,2], ylim=c(min(df[,2:n.thetas]), max(df[,2:n.thetas])),
	 xlab = "div/max(div)", ylab="value of parameter", main="Holonomic ELARS, Diabetes data")
 lines(df[,n.thetas+1], df[,2])
 for (k in (2:p)+1) {
	lines(df[,n.thetas+1], df[,k], type="o")
 }
 abline(v=df[ ,n.thetas+1])
 text(1.02, df[p+1, (1:p + 1)[-7]], (1:p)[-7])
 text(1.02, df[p+1, 8] + 0.0007, 7)
 # text(1.02, df[p+1, 1:p + 1], 1:p)
}

# Gives sequence of thetas going to zero
get.order <- function(df) {
 dims <- dim(df)
 I <- !logical(dims[2])
 I[c(1, dims[2])] <- FALSE
 order <- NULL
 for(i in (dims[1]-1):1) {
	zero <- which(df[i,] == 0 & I)
	order <- c(order, zero)
	I[zero] <- FALSE
 }
 try(print(colnames(X)[order]))
 return(order - 1)
}

############

MLE1 <- mle.full(X.B, Y)
MLE0 <- mle.simplest(X.B, Y)
df <- main(MLE1, MLE0)
make.plot(df)
print(get.order(df))
\end{lstlisting}

%% file: Sections/intro.tex
\thispagestyle{empty}
\section{Introduction} 
\label{sec:introduction}

In model selection, one would ideally want to choose a statistical model that fits the data well, while still being simple enough to allow meaningful interpretations and explanatory power. In this paper we consider model simplification of linear and generalized linear models, where we want to choose a subset of the covariates to include in the model.

In two decades, there have been many advances in sparse modeling. One of the most famous methods is L1-regularization: Least Absolute Shrinkage and Selection Operator (LASSO \cite{LASSO}). LASSO is defined only for the normal linear regression problem. However, the idea of LASSO has been applied to many other problems. For example, Park \& Hastie \cite{park07} considered the generalized linear models and Yuan \& Lin \cite{yuan07} treated the Gaussian graphical models. Least Angle Regression (LARS, \cite{LARS}) is an efficient algorithm for computing the LASSO solution paths. The LARS algorithm is described based on Euclidean geometry because LARS considers the normal linear regression problem. Hirose \& Komaki \cite{hirose2010extension} proposed the ELARS algorithm based on the information geometry of dually flat spaces. ELARS is an algorithm for estimating and selecting parameters in the generalized linear models. The idea of ELARS was applied to edge selection in the Gaussian graphical models \cite{hirose13} and the contingency table models \cite{hirose15}. Another version of geometrical extensions of LARS was given by Augugliaro, Mineo \& Wit \cite{augugliaro13}, and a geometrical approach to sparse modeling was also proposed in \cite{yukawa16}.

The ELARS algorithm by Hirose and Komaki \cite{hirose2010extension} has a computational drawback in that it assumes that the potential function (i.e. the normalizing constant) of the underlying probability distribution function is easy to compute. This is often not the case, which motivates us to use the holonomic gradient method, a computationally efficient method for computing potential functions and their gradients, introduced by Nakayama et al.\ \cite{nakayama2011holonomic}.
A system of linear partial differential equations is called a holonomic system if it has a finite-dimensional space of solutions.
Refer to Section~\ref{sec:hgm} for a more precise description.
If the potential function satisfies a holonomic system, we can use a modification of the holonomic gradient method to keep track of its value at each step of the algorithm and update it when needed in a computationally efficient way.
We call the combined algorithm the holonomic extended LARS algorithm, or HELARS.

The main result of the paper is an implementation in \texttt{R} of the HELARS. We choose the truncated normal distribution as the underlying distribution, as it is simple enough to handle due to its similarities with the well-known normal distribution. Despite the truncated normal having no closed from potential function, our implementation of the algorithm does not use numerical integration.
Of course, the potential function of the truncated normal distribution is nothing but the Gaussian cumulative distribution function, that is implemented as a built-in function in almost all software packages. However, since the truncated normal model is a special case of more complicated models such as the exponential-polynomial distributions \cite{takemurahayakawapolynomial} and the multivariate truncated normal distributions \cite{koyamatakemuraorthant}, our result will become a prototype of the overall method.

The paper is organized as follows. In Section \ref{sec:statistics} we review basic definitions and results concerning generalized linear models. In Section \ref{sec:hgm} we present the holonomic gradient method. Section \ref{sec:bisector_regression} discusses the extended LARS algorithm \cite{hirose2010extension} by Hirose and Komaki, and we look at what necessary changes and additions are needed for the HELARS algorithm. In Section \ref{sec:trunc_example} we use the truncated normal distribution as the underlying distribution, and implement the HELARS algorithm. We validate the algorithm using both real and simulated datasets. Finally, we end with a discussion of the results in Section \ref{sec:discussion}.

%% file: Sections/statistics.tex
\section{Generalized linear models} 
\label{sec:statistics}
In this section we will review some foundations of generalized linear models. We will follow \cite{glm} in our exposition.

\begin{definition}
	Consider a statistical model $\mathcal P = \{p_{\vec \xi} \mid \vec \xi \in \Xi\}$, where $\Xi \subseteq \mathbb{R}^d$. We say that $\mathcal P$ is an exponential family if for $\vec{y} = (y_1,\dotsc,y_n) \in \mathbb{R}^n$, $\vec\xi = (\xi^1,\dotsc, \xi^d)\in \Xi$ we have
\begin{align}
	p_{\vec \xi}(\vec y) = p(\vec y \mid \vec \xi) = e^{C(\vec y) + \vec\xi \cdot \vec F(\vec y) - \psi(\vec\xi)}, \label{eq:gen_exp_fam_statistics_tex}
\end{align}
for some $\vec F \colon \mathbb{R}^n \to \mathbb{R}^d$ and $C \colon \mathbb{R}^n \to \mathbb{R}$, and where $\psi(\vec\xi)$ is the logarithm of the normalizing constant, i.e.
\begin{align*}
	\psi(\vec \xi) = \log \int e^{C(\vec y) + \vec\xi \cdot \vec F(\vec y)} \dd y
\end{align*}
\end{definition}

\begin{example}
	The \emph{normal distribution} is a member of the exponential family. It has the probability density function
	\begin{align*}
		p(y \mid \mu, \sigma^2) &= \frac{1}{\sqrt{2\pi\sigma^2}} \exp \left[ -\frac{(y-\mu)^2}{2\sigma^2} \right]\\
		&=\frac{1}{\sqrt{2\pi \sigma^2}}\exp \left[ -\frac{y^2 - 2y\mu + \mu^2}{2\sigma^2} \right]\\
		&= \frac{1}{\exp(\mu^2/(2\sigma^2))\sqrt{2\pi \sigma^2}}\exp \left[ -\frac{y^2 - 2y\mu}{2\sigma^2} \right].
	\end{align*}
	Note that here the natural parameter is $\vec \xi = \begin{bmatrix}
		-\frac{1}{2\sigma^2} & \frac{\mu}{\sigma^2}
	\end{bmatrix}^T$ and $F(y) = \begin{bmatrix}
		y^2 & y
	\end{bmatrix}^T$. 
\end{example}


\begin{definition}
The \emph{Fisher information matrix} of a distribution $p(\vec y\mid\vec \xi)$ at a point $\vec \xi$ is an $d \times d$ matrix $G(\xi)=(g_{i,j})$ with entries given by
\begin{align*}
	g_{i,j} = E\left[\frac{\partial \log p(\vec y\mid \vec\xi)}{\partial \xi^i}\frac{\partial \log p(\vec y\mid \vec \xi)}{\partial \xi^j} ~\middle|~ \vec \xi \right]
\end{align*}
Equivalently, we may write the elements of the Fisher information matrix as
\begin{align*}
	g_{i,j} = -E\left[\frac{\partial^2}{\partial \xi^i \partial \xi^j} \log p(\vec y\mid \vec\xi) ~\middle|~ \vec\xi \right]
\end{align*}
\label{cor:fisher_as_double_derivative}
\end{definition}

Next we introduce generalized linear models. Assume we have $n$ independent observations $y_1,y_2,\dotsc, y_n$. Each observation $y_i$ is sampled from an exponential family with scalar parameter $\xi_i$, which will depend on a \emph{covariate vector} $\vec{x^i} = (x^i_1, \dotsc, x^i_d)$. The $(n \times d)$ matrix $X = (x^i_j)$ is called the \emph{design matrix}. We assume that the covariate vector $\vec{x_i}$ influences the distribution of $y_i$ only via the \emph{linear predictor} $\eta_i$, defined as
\begin{align*}
	\eta_i := \sum_{j=1}^d \theta_j x^i_j,
\end{align*}
or using matrices $\vec{\eta} := \mtrx{X}\vec{\theta}$,
for some vector $\vec{\theta} \in \mathbb{R}^d$.
We can also add an intercept term by definining a new design matrix $\tilde{\mtrx X} := \begin{bmatrix}
		\mtrx{1_{n\times 1}} & X
	\end{bmatrix}$ so that $\vec{\eta} := \tilde{\mtrx{X}}\vec{\theta},$
for some vector $\vec{\theta} = (\theta_0, \theta_1, \dotsc, \theta_d)^T$. We will always use an intercept term throughout this paper.

The final piece of a generalized linear model, the \emph{link function} $g(\mu)$, determines in which way the linear predictor influences the distribution by setting
\begin{align*}
	\eta_i = g(\mu_i),
\end{align*}
where $\mu_i$ is the expectaion of $y_i$.
The \emph{canonical link} is the link function $g$ for which $\xi_i = \eta_i$ for all $i=1,2,\dotsc,n$.

The combination of an exponential family, design matrix and link function define a \emph{generalized linear model} (GLM). The model has $d+1$ parameters $\theta_0, \theta_1,\dotsc,\theta_d$, that we want to estimate given the response $\vec y$ and design matrix $\mtrx X$. Fitting a GLM is usually more delicate than fitting a linear model. Again, the standard goal is to find the parameters $\theta_0,\theta_1,\dotsc,\theta_d$ which maximize the (log-)likelihood. The log-likelihood of the joint distribution of $n$ observations is $L = \sum_{i=1}^n L_i$,
where $L_i = \log p(y_i \mid \xi_i)$. This is maximized when all the partial derivatives $\frac{\partial L}{\partial \theta_k}$ vanish. In general, the partial derivatives will not be linear functions of $\vec\theta$, so we have to resort to numerical methods to compute the MLE.

A common iterative method to find the estimate $\hat{\vec\theta}$ is the Newton-Raphson method. Note that we will use this method extensively along with the holonomic gradient method in our implementation (see Section \ref{sec:trunc_example}) for both maximum likelihood estimation and other optimization tasks. We start with an initial guess $\vec\theta^{(0)}$. For each $k\geq 0$, we approximate the function at the point $\vec\theta^{(k)}$ with a polynomial of degree 2. Finding the extremum of the approximation is easy, and we set the point reaching the extremum as the next estimate $\vec\theta^{(k+1)}$.

More precisely, let $\vec\theta^{(k)} \in \mathbb{R}^d$ be the current estimate. The Taylor expansion up to the second order term at this point is
\begin{align*}
	\tilde L(\vec\theta) = L(\vec\theta^{(k)}) + \vec u^{(k)} \cdot (\vec \theta - \vec\theta^{(k)}) + \frac{1}{2}(\vec\theta - \vec \theta^{(k)})^T \mtrx H^{(k)} (\vec \theta - \vec \theta^{(k)}),
\end{align*}
where $\vec u^{(k)}$ and $\mtrx H^{(k)}$ are respectively the gradient and Hessian evaluated at $\vec\theta^{(k)}$:
\begin{align*}
	u^{(k)}_i = \left.\frac{\partial L}{\partial \theta_i}\right|_{\vec\theta = \vec\theta^{(k)}} && (\mtrx H^{(k)})_{ij} = \left.\frac{\partial^2 L}{\partial \theta_i\partial \theta_j}\right|_{\vec\theta = \vec\theta^{(k)}}.
\end{align*}
Setting the derivative of $\tilde L$ to zero yields the value of the next estimate
\begin{align*}
	& \vec u^{(k)} + \mtrx H^{(k)} (\vec \theta - \vec \theta^{(k)}) = 0\\
	\implies & \vec\theta^{(k+1)} = \vec\theta^{(k)} - \left( \mtrx H^{(k)} \right)^{-1} \vec u^{(k)}.
\end{align*}
Given a good initial guess, the method will converge to the maximum likelihood estimate as $k\to\infty$.


%% file: Sections/holonomic.tex
\section{Holonomic gradient method} 
\label{sec:hgm}

In this section we will describe the holonomic gradient method (HGM), first proposed by Nakayama et al.\ \cite{nakayama2011holonomic}. Consider first the ``classical'' gradient descent algorithm, which is used to find a local minimum of a function $F \colon \mathbb{R}^n\to \mathbb{R}$. Given a starting point (or initial guess) $\vec{x^{(0)}}$, we know that the value of the function decreases the fastest in the direction opposite to the gradient. In other words, we should choose the next point
\begin{align}
	\vec{x^{(1)}} = \vec{x^{(0)}} - \gamma_1\nabla F(\vec{x^{(0)}}),
\end{align}
for some stepsize $\gamma_1 > 0$. Now, given a suitably chosen stepsize $\gamma_1$, we have $F(\vec{x^{(1)}}) < F(\vec{x^{(0)}})$. We then iterate
\begin{align}\label{eq:iter_k}
	\vec{x^{(k+1)}} = \vec{x^{(k)}} - \gamma_{k+1} \nabla F(\vec{x^{(k)}}),
\end{align}
while choosing a suitable stepsize $\gamma_k > 0$ at each iteration. We can terminate the algorithm when the gradient is small enough (i.e. when we are close to a local minimum), or when a certain number of iterations have elapsed. Details concerning the choice of step size and efficiency of this method will not be discussed here; see for example \cite{num_methods_book}. 

The issue with this method is that it requires computing the gradient $\nabla F(\vec{x^{(k)}})$ at each step. In many statistical applications, the function we want to optimize will be a likelihood function, which will in some cases contain an integral that does not have a closed form expression, and has to be computed using numerical methods. As discussed previously, one such example is the 1-dimensional truncated normal distribution.


The holonomic gradient method takes a different approach to function minimization. The main idea is still the same: we use the same iterative step as in the classical gradient descent
\begin{align}
	\vec{x^{(k+1)}} = \vec{x^{(k)}} - \gamma_{k+1} \nabla F(\vec{x^{(k)}}).
	\label{eqn:step}
\end{align}
The difference is how we compute the gradient. We will construct a vector $\vec{Q}$ and a set of matrices $\mtrx{P_i}$ to form a \emph{Pfaffian system}
\begin{align}
	\frac{\d \vec{Q}}{\d x_i} = \mtrx P_i \vec{Q}.
	\label{eq:intro_pfaffian}
\end{align}
The vector $\vec{Q}$ will be chosen so that the gradient $\nabla F(\vec{x^{(k)}})$ is easily recoverable, typically $\nabla F(\vec{x^{(k)}}) = \mtrx A(\vec{x^{(k)}})\vec{Q}(\vec{x^{(k)}})$ for some matrix $\mtrx A$ with entries in $\mathbb{C}(x_1,\dotsc,x_n)$. With the gradient, we can determine the next point $\vec{x^{(k+1)}}$ using \eqref{eqn:step}.
Given $\vec{Q}(\vec{x^{(k)}})$, the value of $\vec{Q}$ in the previous step, we can compute its value in the next step $\vec{Q}(\vec{x^{(k+1)}})$ by solving the Pfaffian system \eqref{eq:intro_pfaffian} using standard numerical ODE solvers.

Observe that we can also implement the Newton-Raphson method using the holonomic gradient framework. The update step will be
\begin{align}
	\vec{x^{(k+1)}} = \vec{x^{(k)}} - \vec\Hess(F(\vec{x^{(k)}}))^{-1} \vec\nabla F(\vec{x^{(k)}}),
\end{align}
and we can also recover the Hessian easily from the Pfaffian system, since there is a matrix $\mtrx B(\vec{x^{(k)}})$ with elements in $\mathbb{C}(x_1,\dotsc,x_n)$ such that $\vec\Hess(F)(\vec{x^{(k)}}) = \mtrx B(\vec{x^{(k)}}) \vec Q(\vec{x^{(k)}})$.

\subsection{Rings of differential operators}
Let $\mathbb{C}(x_1,\dotsc,x_n)$ denote the ring of rational functions.
\begin{definition}
	The ring of differential operators with rational function coefficients, denoted $R_n$, is
	\begin{align*}
		R_n = \mathbb{C}(x_1,\dotsc,x_n) \langle \partial_1, \dotsc,\partial_n \rangle,
	\end{align*}
	where the operator $\partial_i$ corresponds to differentiation with relation to $x_i$, i.e.
	\begin{align*}
		\partial_i = \frac{\partial}{\partial x_i},
	\end{align*}
	and the operators $x_i$ just multiply by $x_i$.
\end{definition}
Note that the ``multiplication'' operation inside the ring is actually a composition of operators. Since every element in $R_n$ is an operator, there is a natural action $\bullet$ on the set $C^\infty$ of smooth functions. If $f(x_1,\dotsc,x_n) \in C^\infty$, then
\begin{align*}
	\partial_i \bullet f(x_1,\dotsc,x_n) &= \frac{\partial f}{\partial x_i} \\
	x_i \bullet f(x_1,\dotsc,x_n) &= x_i f(x_1,\dotsc,x_n).
\end{align*}
$R_n$ is not a commutative, since
\begin{align*}
	\partial_i x_i = x_i \partial_i + 1,
\end{align*}
because of the chain rule. When $i\neq j$, everything commutes:
\begin{align*}
	\partial_i \partial_j = \partial_j \partial_i\\
	x_i x_j = x_j x_i\\
	x_i \partial_j = \partial_j x_i.
\end{align*}
Because of these commutation rules, any element in $R_n$ can be written as a sum of terms with the $\partial_i$ on the right of each term:
\begin{align*}
	\sum_{\alpha} c_\alpha(x_1,\dotsc,x_n) \partial^\alpha,
\end{align*}
where $\alpha$ is a multi-index, $\partial^\alpha = \partial_1^{\alpha_1}\partial_2^{\alpha_2}\dotsb \partial_n^{\alpha_n}$ and $c_\alpha \in \mathbb{C}(x_1,\dotsc,x_n)$ with only finitely many nonzero $c_\alpha$. Most standard theorems and algorithms from algebraic geometry in the regular polynomial ring $\mathbb{C}[x_1,\dotsc, x_n]$ carry over to $R_n$ with minor modifications.

In particular, Macaulay's theorem will be useful in the next section. Let $\prec$ be a term order on the differential operators $\partial_i$. Let $f = \sum_\alpha c_\alpha \partial^\alpha \in R_n$. The leading term of $f$ is the term $\LT(f) = c_{\alpha'}\partial^{\alpha'}$ such that $\partial^{\alpha} \prec \partial^{\alpha'}$ for all $\alpha \neq \alpha'$ such that $c_\alpha \neq 0$. For an ideal $I \subset R_n$, we define $\LT(I)$ as the set of all leading terms in $I$.

\begin{theorem}[Macaulay's theorem]
Let $I\subset R_n$ be an ideal. The set of \emph{standard monomials}
\begin{align*}
	\{w \in R_n \mid w\ \text{is a monomial},\ w \not\in \LT(I)\}
\end{align*}
is a basis of $R_n/I$ as a vector space over $R_n$.
\end{theorem}
We say an ideal $I\subseteq R_n$ is 0-dimensional if there are finitely many standard monomials. A necessary and sufficient condition for $I$ to be 0-dimensional is that for all $i$ the following holds
\begin{align*}
	I \cap \mathbb{C}(x_1,\dotsc,x_n)\langle \partial_i \rangle \neq \{0\}.
\end{align*}

For more details on computations on rings of differential operators, see \cite{dojo}.

\subsection{Pfaffian systems} 
\label{sub:pfaffian_systems}

Let $f(x_1,\dotsc,x_n) \in C^\infty$ be a function, and $\ell \in R_n$. When $\ell \bullet f = 0$, we say that $f$ is \emph{annihilated} by $\ell$. We say that $f$ is annihilated by an ideal $I \subset R_n$ if $f$ is annihilated by all $\ell \in I$. Observe that if $I = \langle \ell_1,\dotsc,\ell_s \rangle$, $f$ is annihilated by $I$ if and only if $f$ is annihilated by all $\ell_i$. Assume that $I$ is 0-dimensional, and let $s_1 = 1,s_2,\dotsc,s_r$ be the standard monomials, which are the generators of $R_n/I$. For all $1\leq i \leq n$ and $1\leq j \leq r$, we can look at the image of the operator $\partial_i s_j$ in the quotient $R_n/I$ (under the canonical map $p \mapsto p+I$), and write it as a $\mathbb{C}(x_1,\dotsc,x_n)$ linear combination of the basis elements
\begin{align*}
 	\partial_i s_j = \sum_{k=1}^r p^i_{jk}s_k,
\end{align*}
where $p^i_{jk} \in \mathbb{C}(x_1,\dotsc,x_n)$ for all $1 \leq i \leq n$ and $1 \leq j,k \leq r$.
Thus, if we define the vector $\vec S = (s_1, s_2, \dotsc, s_r)^T$, then for each $i$, there is a matrix $\mtrx P_i$ such that
\begin{align}
	\partial_i \vec S = \mtrx P_i \vec S, \label{eq:pfaffian_prelim}
\end{align}
with $(\mtrx P_i)_{jk} = p^i_{jk}$.

Define the vector $\vec Q = (s_1 \bullet f, s_2 \bullet f, \dotsc, s_r \bullet f)^T$. Since $f$ is annihilated by all elements in $I$, Equation \eqref{eq:pfaffian_prelim} is true when we replace $\vec S$ by $\vec Q$. We get the following system of differential equations
\begin{align*}
	\frac{\partial \vec Q}{\partial x_i} = \mtrx P_i \vec Q,
\end{align*}
the Pfaffian system. Because we chose $s_1 = 1$, the gradient of $f$ can be recovered from the first elements of each equation in the Pfaffian system
\begin{align*}
	\vec\nabla f = \begin{bmatrix}
		(\mtrx P_1 \vec Q)_1 \\ (\mtrx P_2 \vec Q)_1 \\ \vdots \\ (\mtrx P_n \vec Q)_1
	\end{bmatrix}
\end{align*}

By following the procedure above, one can construct a Pfaffian system given a 0-dimensional ideal annihilating our function $f$. This is indeed desirable, since finding a 0-dimensional annihilating ideal is often easier than to find a Pfaffian system from scratch. One noteworthy fact is that if $f$ has a holonomic annihilating ideal, then its integral over one variable $\int f \dd x_i$ also has a holonomic annihilating ideal. This is extremely useful when using the holonomic gradient method in maximum likelihood estimation, since the normalizing constant will usually contain an integral. Oaku \cite{oaku_algorithms} describes an algorithm for computing the (holonomic) annihilating ideal of an integral.



%% file: Sections/bisect_regr.tex
\section{Holonomic Extended LARS} 
\label{sec:bisector_regression}
In this section, we will describe the holonomic extended least angle regression algorithm. We will also compute explicit forms for coordinate conversion functions between the e-affine and m-affine coordinates, Fisher information matrix, and divergence for the manifold used in Hirose and Komaki \cite{hirose2010extension}.

Consider a set of observed data $\{y_a, \vec{x^a}=(x^a_1,\dotsc,x^a_d) \mid a=1,2,\dotsc,n\}$, where $\vec{y} = (y_1,\dotsc,y_n)^T$ is the response vector, and $\mtrx X = (x^a_j)$ is the design matrix, which has dimensions $(n \times d)$. We will also add an intercept term to the model, so the design matrix becomes $\tilde{\vec X} = \begin{bmatrix} \mtrx{1_{n\times 1}} & \mtrx X \end{bmatrix}$. We will consider exponential families of the form
\begin{align}
	p(\vec{y} \mid \vec{\xi}) = \exp\left( \sum_{a=1}^n y_a\xi^a + \sum_{b=1}^r u_b(\vec{y})\xi^{b+n} - \psi^*(\vec{\xi}) \right). \label{eq:exp_family_xi}
\end{align}

We will define some notation. Let $\vec\xi$ be the natural parameter, a $n+r$ sized vector containing elements $\xi_i$. We can split $\vec\xi$ into two subvectors: we call $\vec\xi'$ the subvector containing the first $n$ elements, i.e. $\vec\xi' = (\xi^a)_{a=1}^n$, and we call $\vec\xi''$ the subvector containing the last $r$ elements, i.e. $\vec\xi'' = (\xi^b)_{b=n+1}^{n+r}$. Hence $\vec\xi = (\vec\xi', \vec\xi'')^T$. The function $\psi^*(\vec\xi)$ is the potential function of $\vec\xi$, and it is equal to the logarithm of the normalizing constant of the distribution
\begin{align*}
	\psi^*(\vec\xi) = \log \int \exp\left( \sum_{a=1}^n y_a\xi^a + \sum_{b=1}^r u_b(\vec{y})\xi^{b+n} \right) \dd \vec{y}
\end{align*}

In a generalized linear model with canonical link function, the natural parameter is related to linear predictor by	$\vec\xi' = \tilde{\mtrx X} \vec\theta'$, where $\vec\theta' = (\theta_0, \theta_1, \dotsc, \theta_d)^T$ is a parameter vector. In addition, as in \cite{hirose2010extension}, we require $r$ additional parameters that are equal to $\vec \xi''$. Hence we can write $\vec \theta'' = \vec \xi''$, and define $\vec \theta = (\vec\theta', \vec\theta'')$. Equation \eqref{eq:exp_family_xi} thus becomes
\begin{align*}
	p(\vec y \mid \vec\theta) = \exp\left( \vec y^T \tilde{\mtrx X}\vec\theta' + \sum_{b=1}^r u_b(\vec y)\theta^{b+d} - \psi(\vec\theta) \right),
\end{align*}
where the potential function of $\vec\theta$ is $\psi(\vec\theta) = \psi^*(\tilde{\mtrx X} \vec\theta', \vec\theta'')$. Alternatively, define the sufficient statistic
\begin{align*}
 	\vec Y = (y_1 , \dotsc , y_n , u_1(\vec y) , \dotsc , u_r(\vec y))^T,
 \end{align*}
and an $(n+r) \times (d+r+1)$ block-diagonal matrix
\begin{align*}
	\mtrx X_B = \begin{bmatrix}
		\tilde{\mtrx X} & \mtrx{0}_{n\times r} \\
		\mtrx{0}_{r\times (d+1)} & \mtrx{I}_{r\times r},
	\end{bmatrix}
\end{align*}
where $\mtrx{0}_{n\times m}$ and $\mtrx{I}_{n\times n}$ are respectively the $(n\times m)$ zero matrix and the $(n\times n)$ identity matrix. Then we have the identities
\begin{align}
\begin{aligned}
	\vec\xi &= \mtrx X_B \vec \theta \\
	\psi(\vec\theta) &= \psi^*(\mtrx X_B \vec \theta)
	\end{aligned} \label{eq:xi_is_xb_theta}
\end{align}
and the probability density function becomes
\begin{align*}
	p(\vec y \mid \vec \theta) = \exp(\vec Y^T \mtrx X_B \vec \theta - \psi(\vec\theta)).
\end{align*}

The $\vec \xi$ coordinate, being the natural parameter of an exponential family, is the e-affine coordinate of the model manifold. The corresponding m-affine coordinate $\vec\mu$ is the expectation parameter $\vec\mu = \E[\vec Y]$ and its potential function is defined as $\phi^*(\vec \mu) = \vec\xi \cdot \vec\mu - \psi^*(\vec\xi)$. The model manifold defined by the coordinates $\vec\theta$ is a submanifold of the model defined by the $\vec\xi$ coordinates. The e-affine coordinate of this submanifold is $\vec\theta$, and there is a dual m-affine coordinate $\vec\eta$ which is related to $\vec\mu$ by
\begin{align}
	\vec\eta = \E[(\vec Y^T \mtrx X_B)^T] = \mtrx X_B^T \E[\vec Y] = \mtrx X_B^T\vec\mu. \label{eq:eta_is_xbt_mu}
\end{align}

Note that the Fisher information matrix of the model in \eqref{eq:exp_family_xi} is equal to the Hessian of the potential function $\psi^*(\vec\xi)$, denoted $\mtrx G^* = (g^*_{i,j})$. Similarly, denote the Hessian of the potential function of the m-affine coordinates $\phi^*(\mu)$ as the matrix $\mtrx G_* = (g_*^{i,j})$. Since $\vec \xi$ and $\vec \mu$ are dual coordinates, the matrices $\mtrx G^*$ and $G_*$ are inverses of each other.


The Hessian $\mtrx G(\vec\theta) = (g_{i,j})$ of the potential function $\psi(\vec\theta) (= \psi^*(\tilde{\mtrx X}\vec \theta', \vec\theta''))$ can be recovered using the chain rule:
\begin{align}
	g_{i,j} 	= \frac{\partial^2 \psi(\vec\theta)}{\partial \theta^i \partial \theta^j} = \frac{\partial \eta_i}{\partial \theta_j} = \sum_{a=0}^{n+d}\sum_{b=0}^{n+d} \frac{\partial \eta_i}{\partial \mu_a}\frac{\partial \mu_a}{\partial \xi_b}\frac{\partial \xi_b}{\partial \theta_j}. \label{eq:hess_psi_1}
\end{align}
Using the identities in \eqref{eq:xi_is_xb_theta} and \eqref{eq:eta_is_xbt_mu}, we see that
\begin{align*}
	\frac{\partial \eta_i}{\partial \mu_a} &= \frac{\partial (\mtrx X_B^T\vec\mu)_i}{\partial \mu_a} = (\mtrx X_B^T)_{i,a} \\
	\frac{\partial \xi_b}{\partial \theta_j} &= \frac{\partial (\mtrx X_B\vec\theta)_b}{\partial \theta_j} = (\mtrx X_B)_{b,j}.
\end{align*}
Thus \eqref{eq:hess_psi_1} becomes
\begin{align*}
	g_{i,j} 	&= \sum_{a=1}^{n}\sum_{b=1}^{n} (\mtrx X^T_B)_{i,a}g^*_{a,b}(\mtrx X_B)_{b,j},
\end{align*}
which implies that $\mtrx G = \mtrx X_B^T \mtrx G^*\mtrx X_B$. We denote elements of its inverse with superscripts: $\mtrx G^{-1} = (g^{i,j})$. Similar to the previous case, we have
\begin{align}
\begin{aligned}
	(\mtrx G)_{i,j} &= \frac{\partial \eta_i}{\partial \theta^j} = g_{i,j}\\
	(\mtrx G^{-1})_{i,j} &= \frac{\partial \theta^i}{\partial \eta_j} = g^{i,j}
\end{aligned} \label{eq:detadtheta_dthetadeta}
\end{align}

\begin{remark}
	In subsequent sections we will make extensive use of matrix and vector differentiation. The convention in \cite{old_new_matrix_algebra_useful_statistics} will be used: the shape of $\frac{\partial\vec f}{\partial \vec x}$ depends either on the shape of $\vec f$ or the shape of $\vec x^T$. For example, differentiating a scalar by a length $n$ column vector yields a length $n$ row vector
	\begin{align*}
		\frac{\partial f}{\partial \vec x} = \begin{bmatrix}
			\frac{\partial f}{\partial x_1} & \frac{\partial f}{\partial x_2} & \cdots & \frac{\partial f}{\partial x_n}
		\end{bmatrix}
	\end{align*}
	Differentiating a length $m$ column vector by a scalar yields a length $m$ column vector
	\begin{align*}
		\frac{\partial \vec f}{\partial x} = \begin{bmatrix}
			\frac{\partial f_1}{\partial x} & \frac{\partial f_2}{\partial x} & \cdots & \frac{\partial f_m}{\partial x}
		\end{bmatrix}^T
	\end{align*}
	Finally, differentiating a length $m$ column vector by a length $n$ column vector yields an $(m \times n)$ matrix, the Jacobian.
	\begin{align*}
		\mtrx\Jac(\vec f) = \frac{\partial \vec f}{\partial \vec x} = \left(\frac{\partial f_i}{\partial x_j}\right)_{i,j}
	\end{align*}
	This notation allows the natural use of the chain rule for derivatives $
			\frac{\partial \vec f(\vec g(\vec x))}{\partial \vec x} = \frac{\partial \vec f(\vec g(\vec x))}{\partial \vec g(\vec x)}\frac{\partial \vec g(\vec x)}{\partial \vec x}$
	with the usual matrix multiplication between the two terms on the right hand side. Furthermore, we can express the Hessian of a scalar valued function $f(\vec x)$ as $\mtrx\Hess(f) = \frac{\partial^2 f}{\partial \vec x \partial \vec x^T}$

\end{remark}

\subsection{Mixed coordinate conversion} 
\label{sub:mixed_coordinate_conversion}

Next, consider a point $P$ on the dually flat manifold
\begin{align*}
 	S = \{p(\vec y \mid \vec \theta) = \exp(\vec Y^T \mtrx X_B \vec \theta - \psi(\vec\theta))\ \mid\  \vec \theta \in \mathbb{R}^{d+r+1}\}.
\end{align*}It is characterized by the e- and m-affine coordinates $\vec \theta = (\theta^0,\theta^1,\dotsc,\theta^{d+r})$ and $\vec \eta = (\eta_0,\eta_1,\dotsc,\eta_{d+r})$. Alternatively, we may use \emph{mixed coordinates}, i.e. for some $J \subset \{0,1,\dotsc,d+r\}$ we represent $P$ as $(\vec\eta_J, \vec\theta^{\bar{J}})$, where $\vec\eta_J = \{\eta_j \mid j\in J\}$ is the subvector of $\vec\eta$ containing only elements which have indices in $J$, and $\vec\theta^{\bar{J}} = \{\theta^j \mid j\not\in J\}$ is the subvector of $\vec\theta$ containing elements with indices not in $J$.

Let $J \subseteq \{0,1,2,\dotsc, d+r\}$, $\overline{J} = \{0,1,2,\dotsc, d+r\} \setminus J$ and let $P = (\vec\eta_J, \vec\theta^{\bar J})$ denote a mixed coordinate. Essentially, we want to recover $\vec\eta_{\bar J}$ and $\vec\theta^J$ given $\vec\eta_J$ and $\vec\theta^{\bar J}$. Let $\tilde{\vec\theta}(\vec\theta^{J}) = (\vec\theta^J, \vec\theta^{\bar J})$, a function $\mathbb{R}^{|J|} \to \mathbb{R}^{d+r+1}$ obtained by mixing the fixed $\vec\theta^{\bar J}$ and the unknown $\vec\theta^J$ coordinates in the positions defined by $J$. Thus, the function $\tilde{\vec\theta}$ outputs the full $\vec\theta$ coordinates, where $\vec\theta^{\bar J}$ are always constant, and $\vec\theta^J$ are allowed to vary. Similarly, let $\tilde{\vec\eta}(\vec\eta_{\bar J}) = (\vec\eta_J, \vec\eta_{\bar J})$ be the same function for the $\vec\eta$ coordinates. 

We will use Newton's method to find the root of the function
\begin{gather}
	F(\vec\theta^J, \vec\eta_{\bar J}) = \tilde{\vec\eta}(\vec\eta_{\bar J}) - \eta(\tilde{\vec\theta}(\vec\theta^J)). \label{eq:function_mixed2theta_F}
\end{gather}
\begin{proposition}
	The Jacobian $\mtrx{\Jac}(F)$ of $F$ in \eqref{eq:function_mixed2theta_F}, has columns
	\begin{align*}
		(\mtrx{\Jac}(F))_i = \begin{cases} 
						-\left( G(\tilde\theta) \right)_i & \text{ if } i\in J \\
						\vec e_i & \text{ if } i\in \overline J
					\end{cases},
	\end{align*}
	where $G(\tilde\theta)_i$ is the $i$th column of $G(\tilde\theta) = \left. \frac{\partial^2 \psi(\vec\theta)}{\partial \vec\theta \partial\vec\theta^T} \right|_{\vec\theta = \tilde{\vec\theta}}$ (see \eqref{eq:detadtheta_dthetadeta}), and the vector $\vec e_i$ is the $i$th standard basis of $\mathbb{R}^{n+d+1}$. \label{prop:mixed_coord_conv}
\end{proposition}
\begin{proof}
	Let $i\in \overline J$. Then
	\begin{align*}
		\frac{\partial F}{\partial \eta_i} = \frac{\partial}{\partial \eta_i}\tilde{\vec\eta}(\vec\eta_{\bar J}) - 0.
	\end{align*}
	Since the $i$th element of $\tilde{\vec\eta}(\vec\eta_{\bar J})$ is simply $\eta_i$ and none of the other elements depend on $\eta_i$, we have $\frac{\partial F}{\partial \eta_i} = \vec e_i$.

	Next let $i\in J$. Using the chain rule we get
	\begin{align*}
		\frac{\partial F}{\partial \theta^i} &= 0 - \frac{\partial}{\partial \theta^i}\left( \eta(\tilde{\vec\theta}(\vec\theta^J)) \right) \\
		&= - \frac{\partial \eta(\tilde{\vec\theta})}{\partial \tilde{\vec\theta}} \cdot \frac{\partial \tilde{\vec\theta}(\vec\theta^J)}{\theta^i} \\
		&= -(G(\tilde{\vec\theta})) \cdot \vec e_i\\
		&= -(G(\tilde{\vec\theta}))_i
	\end{align*} \qed
\end{proof}

Newton's method will iteratively output a vector $(\theta^J,\eta_{\bar J})$ with the following update step
\begin{align*}
	(\theta^J,\eta_{\bar J})^{(k+1)} = (\theta^J,\eta_{\bar J})^{(k)} - \vec{\Jac}(F)^{-1} F,
\end{align*}
where the Jacobian and $F$ are evaluated at $(\theta^J,\eta_{\bar J})^{(k)}$. Given a suitable initial guess, the method converges very quickly.




\subsection{Extended least angle regression algorithm}
\label{sub:elars_algo}
We describe shortly the algorithm by Hirose and Komaki \cite{hirose2010extension}. Let $I$ be the set containing the indices of covariates present in the model. We first start with the model containing all covariates, that is $I=\{1,2,\dotsc,d\}$, and compute the maximum likelihood estimate $\hat{\vec\theta}_{\mathrm{MLE}}$. In addition, we also compute the maximum likelihood estimate $\hat{\vec\theta^{\emptyset}}$ of the empty model, i.e. the model where $\theta^1,\dotsc,\theta^d = 0$. We will work in the $d$ dimensional submanifold of $S$
\begin{align}
	M = \{\vec\eta \mid \eta_0 = \hat\eta_0^\emptyset, \eta_{d+1} = \hat\eta_{d+1}^\emptyset, \dotsc, \eta_{d+r} = \hat\eta_{d+r}^\emptyset \}, \label{eq:space_m}
\end{align}
and set $\hat{\vec\theta}_{(0)} := \hat{\vec\theta}_{\mathrm{MLE}}$ and $k=1$.

For each $i \in I$, let $\bar{\vec\theta}_i$ be the m-projection of the current point $\hat{\vec\theta}_{(k)}$ to the e-flat submanifold corresponding to $\theta^i = 0$. Let $i^*$ be the coordinate which has smallest divergence between the point $\hat{\vec\theta}_{(k)}$ and its m-projection $\bar{\vec\theta}_i$, and let this divergence be $t^*$. Now for each $i\in I$, look at the m-geodesic connecting $\hat{\vec\theta}_{(k)}$ and $\bar{\vec\theta}_i$, and find the point $\vec\theta^*_i$ along that geodesic that has divergence $t^*$ from $\hat{\vec\theta}_{(k)}$. The estimate for the next step $\hat{\vec\theta}_{(k+1)}$ is constructed as follows: for all $i\in I$, set the $i$th coordinate of $\hat{\vec\theta}_{(k+1)}$ to the $i$th coordinate of $\vec\theta^*_i$, and for all $j\not\in I$, set the $j$th coordinate of $\hat{\vec\theta}_{(k+1)}$ to 0. Notice that the $i^*$th coordinate will also be 0. We now remove $i^*$ from the list of ``active'' covariates $I$, and restart at the beginning of the paragraph, this time in the submanifold
\begin{align*}
	M_I = \{\vec\eta \mid \eta_0 = \hat\eta_0^\emptyset, \eta_{d+1} = \hat\eta_{d+1}^\emptyset, \dotsc, \eta_{d+r} = \hat\eta_{d+r}^\emptyset, \theta^j = 0 ~ \forall j\not\in I \}.
\end{align*}
We quit the algorithm after $d$ steps, when no covariates are left. The divergence funtion used in the submanifold $M_I$, which we will denote by $D^{[I]}$, is the restriction of the KL-divergence on $M$ onto the submanifold. We compute it as follows: denote the coordinates in $M_I$ as $\vec\theta_I = (\theta^i)_{i\in I}$ and $\vec\eta_I = (\eta_i)_{i\in I}$. The potential functions become $\psi_I(\vec\theta_I) = \psi(\vec\theta_I,\vec 0_{\bar J})$ and $\phi_I(\vec\eta_I) = \vec\eta_I \cdot \vec\theta_I - \psi_I(\vec\theta_I)$. The divergence is then
\begin{align*}
	D^{[I]}(p,q) = \phi_I(\vec\eta_I(p)) + \psi_I(\vec\theta_I(q)) - \vec\eta_I(p) \cdot \vec\theta_I(q).
\end{align*}

The algorithm starts from the full model and proceeds step by step towards the empty model, which is the opposite direction compared to the LARS. Other than that, the geometric idea of the algorithm is the same as in LARS: at each step $k$, we move the current estimate $\hat{\vec\theta}^{(k)}$ towards the origin, in a direction that bisects the m-geodesics corresponding to each m-projection. We hit the next estimate $\hat{\vec\theta}^{(k+1)}$ exactly when the first of the coordinates $i\in I$ of the vector $\hat{\vec\theta}^{(k)}$ hits 0.

The following pseudocode describes the algorithm more precisely. The algorithm is described in the submanifold $M$ of \eqref{eq:space_m}, so we do not write down coordinates $0, d+1, \dotsc, d+r$ explicitly. We input the data (observations and design matrix) and an underlying distribution (essentially the functions $u_1(\vec y), \dotsc, u_r(\vec y)$ in \eqref{eq:exp_family_xi}), and we get as an output a sequence of estimators $(\hat{\vec\theta}_{(0)}, \dotsc, \hat{\vec\theta}_{(d)})$, where the estimator obtained in the $k$th step corresponds to a model with $k$ covariates removed.
\begin{enumerate}
	\item Let $I = \{1,2,\dotsc, d\}$, $\hat{\vec\theta}_{(0)} := \hat{\vec\theta}_{\mathrm{MLE}}$, and $k=0$.
	
	\item For all $i \in I$, let $M(i,0,I) = \{\vec\theta \mid \theta^i = 0, \theta^j = 0~(j\not\in I)\} = M(I\setminus \{i\})$ and calculate the m-projection $\overline{\vec\theta}(i,I)$ of $\hat{\vec \theta}_{(k)}$ on $M(i,0,I)$.
	
	\item Let $t^* = \min_{i\in I} D^{[I]}(\hat{\vec\theta}_{(k)}, \overline{\vec\theta}(i,I))$ and $i^*=\arg\min_{i\in I} D^{[I]}(\hat{\vec\theta}_{(k)}, \overline{\vec\theta}(i,I))$.
	
	\item For every $\alpha^i \in \mathbb{R}, i\in I$, let $M(i,\alpha^i, I) = \{\theta \mid \theta^i = \alpha^i, \theta^j = 0~(j\not\in I)\}$. For every $i \in I$, compute $\alpha^i$ such that the m-projection $\overline{\vec\theta}'(i,\alpha^i,I)$ of $\hat{\vec\theta}_{(k)}$ on $M(i, \alpha^i, I)$ satisfies $t^* = D^{[I]}(\hat{\vec\theta}_{(k)}, \overline{\vec\theta}'(i,\alpha^i,I))$.

	\item Let $\hat\theta^i_{(k+1)} = \alpha^i ~ (i\in I)$ and $\hat\theta^j_{(k+1)} = 0 ~ (j\not\in I)$.

	\item If $k=d-1$, then go to step 7. If $k<d-1$, then go to step 2 with $k := k+1$, $I := I\setminus \{i^*\}$.

	\item Let $\hat{\vec\theta}_{(d)} = 0$. Output $\hat{\vec\theta}_{(0)},\dotsc,\hat{\vec\theta}_{(d)}$ and quit the algorithm.
\end{enumerate}

We can now rank the covariates in order of importance by looking at the output of the algorithm. The zeroth estimator $\hat{\vec\theta}_{(0)}$ was defined as the maximum likelihood estimate of the full model containing every covariate, and at each subsequent estimator, one of the components will vanish, i.e. the $k$th estimator $\hat{\vec\theta}_{(k)}$ will have exactly $k$ of its elements equal to zero. The element that vanishes corresponds to the covariate that is deemed the least impactful at that particular step. Thus by looking at the order in which the covariates vanish in the sequence $\hat{\vec\theta}_{(0)},\dotsc,\hat{\vec\theta}_{(d)}$, we can order the covariates from least to most important.

\subsection{Adding holonomicity} 
\label{sub:adding_holonomicity}
We will focus our attention to the potential function $\psi(\vec\xi)$ in \eqref{eq:exp_family_xi}, which can be written as
\begin{align*}
	\psi^*(\vec\xi) = \log \int \exp\left( \sum_{a=1}^n y_a\xi^a + \sum_{b=1}^r u_b(\vec{y})\xi^{b+n} \right) \dd \vec{y}.
\end{align*}
Whether or not $\psi^*(\vec\xi)$ has a closed form representation depends on the underlying distribution. We could use numerical integration if no closed form expression for $\psi^*(\vec\xi)$ exists, but such an approach is computationally inefficient.

Instead, we assume that the potential function can be written as
\begin{align}
	\psi^*(\vec\xi) = G_\psi(\vec\xi, \vec L(\vec\xi)), \label{eq:adding_holonicity}
\end{align}
where $G_\psi$ is an elementary function with easily computable derivatives, and $\vec L(\vec\xi)$ is a scalar or vector valued function with a set of Pfaffian systems	$\frac{\partial L_i}{\partial \xi_j} = \mtrx H(\vec\xi, \vec L(\vec\xi))_{i,j}$, or using matrix notation
\begin{align*}
	\frac{\partial \mtrx L}{\partial \vec \xi} = \mtrx H(\vec \xi, \vec L(\vec \xi)).
\end{align*}
Now we obtain the gradient of $\psi^*(\vec\xi)$ as a function of $\vec \xi$ and $\vec L(\vec\xi)$
\begin{align}
	\frac{\partial \psi^*}{\partial \vec \xi} = \frac{\partial G_\psi}{\partial \vec\xi} + \frac{\partial G_\psi}{\partial \vec L} \frac{\partial \vec L}{\partial \vec\xi}. \label{eq:psistar_is_pfaffian}
\end{align}
The derivatives $\frac{\partial G_\psi}{\partial \vec\xi}$ and $\frac{\partial G_\psi}{\partial \vec L}$ are easily computed since we are assuming that $G_\psi$ is an elementary function. We can also write the Fisher information matrix, which is equal to the Hessian of $\psi^*(\vec\xi)$, by differentiating \eqref{eq:psistar_is_pfaffian}. 

\begin{example}[Truncated normal]
	When each observation is distributed according to the truncated normal distribution, we get a potential function of the form
	\begin{align*}
		\psi^*(\vec\xi) = \sum_{a=1}^n \log \int_0^\infty \exp\left( y\xi^a + y^2\xi^{n+1} \right) \dd y
	\end{align*}
	If we define
	\begin{align*}
		\vec L(\vec\xi) = \begin{bmatrix}
			\log \int_0^\infty \exp(y \xi^1 + y^2 \xi^{n+1}) \dd y \\ \log \int_0^\infty \exp(y \xi^2 + y^2 \xi^{n+1}) \dd y \\ \vdots \\ \log \int_0^\infty \exp(y \xi^n + y^2 \xi^{n+1}) \dd y
		\end{bmatrix},
\end{align*}
	then $G_\psi(\vec\xi, \vec L(\vec\xi)) = \sum_{a=1}^n L(\vec\xi)_a = \psi^*(\vec\xi)$. By \eqref{eq:pfaffian_of_L_a_xi} $\vec L$ has a Pfaffian system. See Section \ref{sec:trunc_example} for complete details. \label{example:trunc_normal_L}
\end{example}

\subsubsection{Holonomic update of the vector $L$} 
\label{sub:holonomic_update_of_the_vector_l_}
Nearly every step of the algorithm requires the knowledge of the vector $L$ at some point $P$ with coordinates $\vec\xi$. For example in the case of the truncated normal distribution in Example \ref{example:trunc_normal_L}, computing the vector $\vec L$ requires $n$ separate numerical integrations. 
Using numerical methods to compute $L(\vec\xi)$ at every step is computationally costly. Fortunately we have a Pfaffian system for every element $L_a(\vec\xi)$ in \eqref{eq:pfaffian_of_L_a_xi}. Given another point $\vec\xi_{\mathrm{old}}$ and $\vec L(\vec\xi_{\mathrm{old}})$, we can use standard ODE solvers such as Runge-Kutta to find the value of $\vec L(\vec\xi)$ at some other point $\vec\xi$. In the implementation we use the \texttt{R} package \texttt{hgm} \cite{R_hgm_package} by Takayama et al.\ , which uses the RK4(5)7 method from Dormand and Prince \cite{rk54dp7}.

We can also find the value of $\vec L$ after a change in $\vec\theta$ coordinates. Since $\vec\xi = \mtrx X_B \vec \theta$ we have
\begin{align*}
	\frac{\partial \vec L(\vec\xi)}{\partial \vec \theta} = \frac{\partial \vec L(\vec\xi)}{\partial \vec\xi} \frac{\partial \mtrx X_B \vec \theta}{\partial\vec \theta} = \frac{\partial \vec L(\vec\xi)}{\partial \vec\xi}\cdot \mtrx X_B
\end{align*}
Again, if both $\vec\theta_{\mathrm{old}}$ and $\vec L(\vec\theta_{\mathrm{old}})$ are known, then we can use numerical ODE solvers to obtain $\vec L(\vec\theta)$.

There are also cases where we need to conduct the holonomic update step in terms of mixed coordinates. Let $J \subset \{0, 1, \dotsc, d+r\}$ and assume $P_\mathrm{old} = (\vec\theta^{\bar J}_\mathrm{old}, {\vec\eta_J}_\mathrm{old})$ and $\vec L(P_\mathrm{old})$ are known. In order to obtain $\vec L(P)$ for some other $P = (\vec\theta^{\bar J}, \vec\eta_J)$, we will need to find a Pfaffian system for $\vec L$ in terms of the mixed coordinates
\begin{theorem}
	Let $\emptyset \neq J \subsetneq \{0, 1, \dotsc, d+r\}$ be a nonempty, strict subset and let $\vec\rho = (\vec\theta^{\bar J}, \vec\eta_J)$ denote a mixed coordinate. For a vector $\vec \rho$ with $d+r+1$ elements, let $\vec \rho_J$ denote the subvector $(\rho_j)_{j\in J}$, and similarly $\vec \rho_{\bar J} = (\rho_j)_{j\not\in J}$. Let
	\begin{align*}
	 	\fullfunction{\vec\theta^*}{\mathbb{R}^{d+r+1}}{\mathbb{R}^{d+r+1}}{\vec\rho}{\vec\theta}
	 \end{align*} be the function that maps the mixed coordinates $\vec\rho$ to the $\vec\theta$ coordinate. 
	 Then
	\begin{align*}
		\frac{\partial \vec L}{\partial \vec\rho} = \frac{\partial \vec L}{\partial \vec \theta} \frac{\partial \vec\theta^*}{\partial \vec \rho},
	\end{align*} where
	\begin{align*}
	 	\frac{\partial \vec \theta^*_J}{\partial \vec \rho_J} &= \left( \frac{\partial \vec\eta_J}{\partial\vec\theta^J} \right)^{-1} &&&
	 	\frac{\partial \vec \theta^*_J}{\partial \vec \rho_{\bar J}} &= \left( \frac{\partial \vec\eta_J}{\partial\vec\theta^J} \right)^{-1} \cdot \left(- \frac{\partial \vec\eta_J}{\partial\vec\theta^{\bar J}}\right) \\
	 	\frac{\partial \vec \theta^*_{\bar J}}{\partial \vec \rho_J} &= \mtrx 0_{|\bar J| \times |J|} &&&
	 	\frac{\partial \vec \theta^*_{\bar J}}{\partial \vec \rho_{\bar J}} &= \mtrx I_{|\bar J| \times |\bar J|}
	 \end{align*} 
	 Furthermore, $\frac{\partial \vec L}{\partial \vec \rho}$ is a function of $\vec\rho$ and $\vec L(\vec\rho)$. \label{thm:holonomic_mixed_update}
\end{theorem}
\begin{proof}
	We wish to find the derivative of $\vec L$ in terms of some mixed coordinates $\vec\rho$. We will first convert the mixed coordinates $\vec\rho$ to $\vec\theta$ coordinates, and then evaluate the derivative. By the chain rule, we obtain the first part of the theorem
	\begin{align*}
		\frac{\partial \vec L(\vec\theta^*(\vec\rho))}{\partial\vec\rho} = \frac{\partial \vec L}{\partial\vec\theta}\frac{\partial\vec\theta^*}{\partial\vec\rho}.
	\end{align*}

	By definition, $\vec\rho_J = \eta_J$ and $\vec\rho_{\bar J} = \theta^{\bar J}$, and the function $\vec\theta^*$ satisfies the following identities
	\begin{gather}
		\vec\theta^*(\vec\rho)_{\bar J} = \vec\theta^{\bar J} \label{pf:eq:theta_rho}\\
		\vec\eta(\vec\theta^*(\vec\rho))_J = \vec\eta_J \label{pf:eq:eta_rho}
	\end{gather}

	Let $i \in \bar J$ and $j \in J$. Then clearly
	\begin{align*}
		\frac{\partial\theta^*_i}{\partial\rho_j} = \frac{\partial\theta^i}{\partial\eta_j} = 0,
	\end{align*}
	since the components of $\vec \rho$ do not depend on each other. Likewise, if both $i,j \in \bar J$, then
	\begin{align*}
		\frac{\partial \theta^*_i}{\partial\rho_j} = \frac{\partial \theta^i}{\partial \theta^j} = \delta_i^j.
	\end{align*}
	Hence we get $\frac{\partial \vec \theta^*_{\bar J}}{\partial \vec \rho_J} = \mtrx 0_{|\bar J| \times |J|}$ and $\frac{\partial \vec \theta^*_{\bar J}}{\partial \vec \rho_{\bar J}} = \mtrx I_{\bar J \times \bar J}$.

	Differentiating both sides of \eqref{pf:eq:eta_rho} by $\vec\rho_J$, we have
	\begin{align*}
		\frac{\partial \vec\eta(\vec\theta^*(\vec\rho))_J}{\partial \vec\rho_J} = \frac{\partial \vec\eta_J}{\partial \vec\rho_J},
	\end{align*}
	where the right-hand side becomes $\frac{\partial \vec\eta_J}{\partial \vec\eta_J} = \mtrx I$, and the left-hand side becomes
	\begin{align*}
		\frac{\partial \vec\eta(\vec\theta^*(\vec\rho))_J}{\partial \vec\rho_J} = \frac{\partial \vec\eta(\vec\theta^*(\vec\rho))_J}{\partial \vec\theta^*(\vec\rho)}\frac{\partial \vec\theta^*(\vec\rho)}{\partial \vec\rho_J} = \frac{\partial \vec\eta_J}{\partial \vec\theta^J}\frac{\partial \vec\theta^*(\vec\rho)_J}{\partial \vec\rho_J},
	\end{align*}
	since $\frac{\partial \vec \theta^*_{\bar J}}{\partial \vec \rho_J} = \mtrx 0$. Thus
	\begin{align*}
		\frac{\partial \vec\eta_J}{\partial \vec\theta^J}\frac{\partial \vec\theta^*(\vec\rho)_J}{\partial \vec\rho_J} = \vec I \implies \frac{\partial \vec\theta^*(\vec\rho)_J}{\partial \vec\rho_J} = \left( \frac{\partial \vec\eta_J}{\partial \vec\theta^J} \right)^{-1}
	\end{align*}

	Finally, differentiate both sides of \eqref{pf:eq:eta_rho} by $\vec\rho_{\bar J}$ to get
	\begin{align*}
		\frac{\partial \vec\eta(\vec\theta^*(\vec\rho))_J}{\partial\vec\rho_{\bar J}} = \frac{\partial\vec\eta_J}{\partial\vec\rho_{\bar J}}.
	\end{align*}
	The right hand side is equal to $\frac{\partial\vec\eta_J}{\partial\vec\theta^{\bar J}} = 0$, since once again the elements of $\vec\rho$ do not depend on each other. The left-hand side becomes
	\begin{align*}
		\frac{\partial \vec\eta(\vec\theta^*(\vec\rho))_J}{\partial\vec\theta^*(\vec\rho)} \frac{\partial \vec\theta^*(\vec\rho)}{\partial\vec\rho_{\bar J}} &= \frac{\partial \vec\eta(\vec\theta^*(\vec\rho))_J}{\partial\vec\theta^*(\vec\rho)_J} \frac{\partial \vec\theta^*(\vec\rho)_J}{\partial\vec\rho_{\bar J}} + \frac{\partial \vec\eta(\vec\theta^*(\vec\rho))_J}{\partial\vec\theta^*(\vec\rho)_{\bar J}} \frac{\partial \vec\theta^*(\vec\rho)_{\bar J}}{\partial\vec\rho_{\bar J}}\\[1em]
		&= \frac{\partial \vec\eta_J}{\partial\vec\theta_J} \frac{\partial \vec\theta^*(\vec\rho)_J}{\partial\vec\rho_{\bar J}} + \frac{\partial \vec\eta_J}{\partial\vec\theta_{\bar J}} \frac{\partial \vec\theta^{\bar J}}{\partial\vec\theta^{\bar J}}\\
		&= \frac{\partial \vec\eta_J}{\partial\vec\theta_J} \frac{\partial \vec\theta^*(\vec\rho)_J}{\partial\vec\rho_{\bar J}} + \frac{\partial \vec\eta_J}{\partial\vec\theta_{\bar J}} \mtrx I.
	\end{align*}
	Hence
	\begin{align*}
		\frac{\partial \vec\eta_J}{\partial\vec\theta_J} \frac{\partial \vec\theta^*(\vec\rho)_J}{\partial\vec\rho_{\bar J}} + \frac{\partial \vec\eta_J}{\partial\vec\theta_{\bar J}} = 0 \implies \frac{\partial \vec\theta^*(\vec\rho)_J}{\partial\vec\rho_{\bar J}} = \left( \frac{\partial \vec\eta_J}{\partial\vec\theta_J} \right)^{-1} \left( - \frac{\partial \vec\eta_J}{\partial\vec\theta_{\bar J}}\right)
	\end{align*}

	Finally, $\frac{\partial \vec L}{\partial \vec \rho}$ is indeed a function of $\vec\rho$ and $\vec L(\vec\rho)$, since $\frac{\partial \vec L}{\partial \vec\theta}$, $\frac{\partial \vec\eta}{\partial\vec\theta} = \frac{\partial^2 \psi(\vec\theta)}{\partial \vec\theta \partial\vec\theta^T}$ is a function of $\vec\rho$ and $\vec L(\vec\rho)$\footnote{after appropriate coordinate conversions.} based on the discussion in the beginning of Subsection \ref{sub:adding_holonomicity}. \qed
\end{proof}
\subsubsection{Holonomic m-projections} 
\label{sub:holonomic_m_projections}
Using the results of Theorem \ref{thm:holonomic_mixed_update} we can now carry out m-projections and recover the vector $\vec L$ at the projected point given the value of $\vec L$ at the previous point. Let $\emptyset \neq I \subset \{0,1,\dotsc,d+r\}$, $i\not\in I$ and $\alpha \in \mathbb{R}$. In the algorithm, all of the m-projections will be to the space $M(i,\alpha,I) = \{\theta \mid \theta^i = \alpha, \theta^j = 0~(j\not\in I)\}$. Let a point $P$ have the dual coordinates $\vec\theta$ and $\vec\eta$. The m-projection of $P$ onto $M(i,\alpha,I)$ will have the mixed coordinates $\vec\rho = (\vec\theta^I, \theta^i = \alpha, \vec\eta_{\bar I \setminus \{i\}})$.
 In other words, we first convert the point $P$ to mixed coordinates according to the set $I \cup \{i\}$ to get $\vec\rho_0 = (\vec\theta^{I\cup \{i\}}, \vec\eta_{\overline{I\cup\{i\}}})$, and then send the element $\theta^i$ to $\alpha$ to get $\vec\rho$. Given $\vec L(\vec\rho_0)$ ($= \vec L(P)$) and the Pfaffian system, we may now use Theorem \ref{thm:holonomic_mixed_update} to obtain $L(\vec\rho)$, and thus recover the full $\vec\theta$ coordinates from the mixed coordinates.

\subsection{Holonomic extended LARS algorithm}
\label{sub:helars_algo}
The holonomic extended LARS algorithm is our main result. The algorithm works exactly as the extended LARS algorithm described in Subsection \ref{sub:elars_algo}, but now we have also to specify a Pfaffian system for $\vec L(\vec\xi)$ as an input in addition to the data (response $\vec y$ and design matrix $\mtrx X$) and the underlying distribution ($u_1(\vec y),\dotsc,u_r(\vec y)$). Again, we describe the algorithm in the submanifold $M \subset S$ (see \eqref{eq:space_m}), so we will mostly ignore the coordinates indexed by $0, d+1, \dotsc, d+r$ in the vectors $\vec\mu$ and $\vec\theta$. We will only compute them at the end of step 5, because they are needed for the initial guesses of the numerical solvers.

We get as an output the a sequence of estimators $\hat{\vec\theta}_{(0)},\dotsc,\hat{\vec\theta}_{(d)}$, where the $k$th estimator $\hat{\vec\theta}_{(k)}$ corresponds to a model with $d-k$ covariates. The holonomic extended bisector regression algorithm thus looks as follows
\begin{enumerate}
	\item Let $I = \{1,2,\dotsc, d\}$, $\hat{\vec\theta}_{(0)} := \hat{\vec\theta}_{\mathrm{MLE}}$, and $k=0$. Compute $\vec L(\hat{\vec\theta}_{(0)})$.
	
	\item For all $i \in I$, let $M(i,0,I) = \{\vec\theta \mid \theta^i = 0, \theta^j = 0~(j\not\in I)\} = M(I\setminus \{i\})$ and calculate the holonomic m-projection $\overline{\vec\theta}(i,I)$ of $\hat{ \vec\theta}_{(k)}$ on $M(i,0,I)$ and obtain the vector $\vec L(\overline{\vec\theta}(i,I))$.
	
	\item Let $t^* = \min_{i\in I} D^{[I]}(\hat{\vec\theta}_{(k)}, \overline{\vec\theta}(i,I))$ and $i^*=\arg\min_{i\in I} D^{[I]}(\hat{\vec\theta}_{(k)}, \overline{\vec\theta}(i,I))$.
	
	\item For any $\alpha^i \in \mathbb{R}, i\in I$, let $M(i,\alpha^i, I) = \{\vec\theta \mid \theta^i = \alpha^i, \theta^j = 0~(j\not\in I)\}$. For every $i \in I$, compute $\alpha^i$ such that the m-projection $\overline{\vec\theta}'(i,\alpha^i,I)$ of $\hat{\vec\theta}_{(k)}$ on $M(i, \alpha^i, I)$ satisfies $t^* = D^{[I]}(\hat{\vec\theta}_{(k)}, \overline{\vec\theta}'(i,\alpha^i,I))$.

{
	\item Let $\hat\theta^i_{(k+1)} = \alpha^i ~ (i\in I)$ and $\hat\theta^j_{(k+1)} = 0 ~ (j\not\in I)$.%
    \def\OldComma{,}
    \catcode`\,=13
    \def,{%
      \ifmmode%
        \OldComma\discretionary{}{}{}%
      \else%
        \OldComma%
      \fi%
    }%
    The $(k+1)$th estimate will have mixed coordinates $\hat{\vec\rho}_{(k+1)} = (\hat\eta_0^\emptyset, \hat\theta^1_{(k+1)}, \dotsc, \hat\theta^d_{(k+1)}, \hat\eta_{d+1}^\emptyset,\dotsc,\hat\eta_{d+r}^\emptyset)$. Use the holonomic update to compute $\vec L(\hat{\vec\rho}_{(k+1)})$ using the value $\vec L(\hat{\vec\rho}_{(k)})$, and then use this to obtain the remaining coordinates $\hat{\theta}_{(k+1)}^0, \hat{\theta}_{(k+1)}^{d+1}, \dotsc, \hat{\theta}_{(k+1)}^{d+r}$.
  }

	\item If $k=d-1$, then go to step 7. If $k<d-1$, then go to step 2 with $k := k+1$, $I := I\setminus \{i^*\}$.

	\item Let $\hat{\vec\theta}_{(d)} = 0$. Output $\hat{\vec\theta}_{(0)},\dotsc,\hat{\vec\theta}_{(d)}$ and quit the algorithm.
\end{enumerate}

As in Subsection \ref{sub:elars_algo}, by looking at the order in which the covariates vanish in the sequence $\hat{\vec\theta}_{(0)},\dotsc,\hat{\vec\theta}_{(d)}$, we can determine the order of importance of the covariates.


%% file: Sections/truncated_normal.tex
\section{A worked out example: the truncated normal distribution} 
\label{sec:trunc_example}

In this section we will work out the implementation of the Holonomic Extended Least Angle Regression algorithm with the truncated normal distribution. The algorithm was implemented in the R programming language \cite{R}, and the code can be found in \cite{implementation}.

\subsection{Introduction} 
\label{sub:trunc_norm_introduction}
The truncated normal distribution is defined as the restriction of the normal distribution to the positive real axis. Its probability density function is
\begin{align*}
	p(y \mid \mu, \sigma^2) = \frac{e^{-(y-\mu)^2/(2\sigma^2)}}{\int_0^\infty e^{-(y-\mu)^2/(2\sigma^2)} \dd y},
\end{align*}
where $y\in(0,\infty)$, $\mu \in \mathbb{R}$ and $\sigma \in (0,\infty)$. By expanding, we can write the probability density function as a function of natural parameters
\begin{align}
	p(y \mid \xi_1, \xi_2) = \frac{e^{\xi_1 y + \xi_2 y^2}}{\int_0^\infty e^{\xi_1 y + \xi_2 y^2} \dd y}, \label{eq:trunc_norm_natural_single}
\end{align}
where $\xi_1 = \frac{\mu}{\sigma^2}$ and $\xi_2 = -\frac{1}{2\sigma^2}$. From the form above we see that the truncated normal distribution belongs to the exponential family. Note also that the normalizing constant $A(\xi_1, \xi_2) = \int_0^\infty e^{\xi_1 y + \xi_2 y^2} \dd y$ does not in general have a closed form, and converges if and only if $\xi_2 < 0$. A generalization of the truncated normal distribution are the \emph{exponential-polynomial distributions}, of the form
\begin{align*}
	f(y \mid \xi_1,\dotsc, \xi_d) = \frac{\exp(\xi_n y^n + \dotsb + \xi_1 y)}{\int_0^\infty \exp(\xi_n y^n + \dotsb + \xi_1 y) \dd y}
\end{align*}
for $y>0$ and $\xi_n < 0$. This family of distributions and their usage with the holonomic gradient method has been studied in Hayakawa and Takemura \cite{takemurahayakawapolynomial}.

We can naturally construct a generalized linear model using the canonical link where each observation is distributed according to the truncated normal distribution. Given a sample $\vec{y} = (y_1,\dotsc,y_n)$, assume that each $y_i$ is independent and distributed according to a truncated normal distribution with a unique mean parameter $\mu_i$ and a common variance parameter $\sigma^2$. Hence, using the notation in equation \eqref{eq:trunc_norm_natural_single} each observation has their own $\xi_1$ parameter, and $\xi_2$ is the same in each observation. To make the notation consistent with \cite{hirose2010extension}, for each $i=1,\dotsc,n$, the ``$\xi_1$ parameter'' of observation $i$ will be called $\xi^i$ and the common ``$\xi_2$ parameter'' will be called $\xi^{n+1}$. With this notation, each observation will have the distribution
\begin{align*}
	p(y_i \mid \xi^i, \xi^{n+1}) = \frac{e^{\xi^i y_i + \xi^{n+1}y_i^2}}{A(\xi^i, \xi^{n+1})},
\end{align*}
and since every observation is independent, the joint distribution of $\vec{y}$ is
\begin{align*}
	p(y_1,\dotsc, y_n \mid \xi^1, \dotsc, \xi^n, \xi^{n+1}) = \frac{e^{\sum_{a=1}^n \xi^ay_a + \left( \sum_{a=1}^n y_a^2 \right) \xi^{n+1}}}{\prod_{a=1}^n A(\xi^a, \xi^{n+1})}
\end{align*}

In the generalized linear model, each observation $y_i$ is explained by a set of $d$ explanatory variables $x^i_1, \dotsc, x^i_d$. With the canonical link function in particular, the natural parameter is simply an affine combination of the explanatory variables, i.e. for some real numbers $\theta^0, \theta^1, \dotsc, \theta^d$, we have $\xi^i = \theta^0 + \theta^1 x^i_1 + \dotsm + \theta^d x^i_d$.

We will now define some notation. Let $\mtrx X = (x^i_j)$ be the $(n \times d)$ design matrix, $\vec{\theta}' = (\theta^0, \theta^1, \dotsc, \theta^d)^T$ and $\vec{\xi}' = (\xi^1, \dotsc, \xi^n)^T$. If $\tilde{\mtrx X} = \begin{bmatrix}
	\vec{1_n} & \mtrx X
\end{bmatrix}$, where $\vec{1_n}$ is a column vector of size $n$ where each element is $1$, then we have $\vec{\xi}' = \tilde{\mtrx{X}} \vec\theta'$. We can also define a block-diagonal matrix $\mtrx{X}_B$ and vector $\vec{Y}$ as
\begin{align*}
	\mtrx{X}_B = \begin{bmatrix}
		\tilde{\mtrx{X}} & \vec{0} \\ \vec{0} & 1
	\end{bmatrix} && \vec{Y} = \begin{bmatrix}
		y_1 \\ y_2 \\ \vdots \\ y_n \\ \sum_{a=0}^n y_a^2
	\end{bmatrix}
\end{align*}
As we defined in Section \ref{sec:bisector_regression}, we have $\theta^{n+1} = \xi^{n+1}$. If we set $\vec{\theta} = (\theta^0, \dotsc, \theta^{d+1})^T$ and $\vec{\xi} = (\xi^1,\dotsc,\xi^{n+1})^T$ we have $\vec\xi = \mtrx{X}_B\vec\theta$, and we can write the pdf of the model as
\begin{align}
	p(\vec y \mid \vec \theta) = \frac{\exp(\vec{Y}^T \mtrx{X}_B \vec \theta)}{\prod_{a=1}^n A_a(\mtrx{X}_B \vec \theta)} = \exp(\vec{Y}^T \mtrx{X}_B \vec \theta - \psi(\vec\theta)), ~\text{ when } \theta^{d+1} < 0\label{eq:pdf_trunc_norm_theta}
\end{align}
where $A_a(\vec \xi) = A(\xi^a, \xi^{n+1}) = \int_0^\infty \exp(\xi^a y + \xi^{n+1} y^2) \dd y $ is the normalizing constant of the $a$th observation, and $\psi(\vec\theta) = \psi^*(\mtrx{X}_B \vec \theta) = \sum_{a=1}^n \log A_a(\mtrx{X}_B\vec\theta)$ is the potential function. 

\subsection{Normalizing constant as a holonomic system} 
\label{sub:normalizing_constant_as_a_holonomic_system}
Next we construct a holonomic system for the normalizing constant. We denote the differential operators by the symbol $\partial$ with the appropriate subscript. For example, we denote $\frac{\partial}{\partial \xi^{n+1}}$ by $\partial_{\xi^{n+1}}$. We will also omit the symbol $\bullet$ used to denote the application of an operator to a function when its usage is clear from context. In addition, any subscript or superscript $a$ will take integer values in $[1,n]$.

We start by looking at the function
\begin{align*}
	A_a(\vec\xi) = A(\xi^a, \xi^{n+1}) = \int_0^\infty \exp(\xi^a y + \xi^{n+1} y^2) \dd y,
\end{align*}
which is defined when $\xi^{n+1} < 0$. Any partial derivative of $A_a$ can be expressed as a partial derivative in terms of $\xi^a$, since
\begin{align}
	\partial_{\xi^{n+1}} A_a = \int_0^\infty y^2 \exp(\xi^a y + \xi^{n+1} y^2) \dd y = \partial_{\xi^a}^2 A_a. \label{eq:deriv_a_xi_nplus1_xi_n}
\end{align}
Furthermore, we can use integration by parts on $A_a$ to get
\begin{align*}
	A_a =& \int_0^\infty e^{\xi^a y}e^{\xi^{n+1}y^2} \dd y\\
	=& \frac{1}{\xi^a} \left[ e^{\xi^ay + \xi^{n+1} y^2} \right]_0^\infty - \frac{2\xi^{n+1}}{\xi^a} \int_0^\infty ye^{\xi^ay + \xi^{n+1}y^2} \dd y \\
	=& -\frac{1}{\xi^a} - \frac{2\xi^{n+1}}{\xi^a} \partial_{\xi^a} A_a,
\end{align*}
and hence the following partial differential equation holds
\begin{align}
	(\xi^a + 2\xi^{n+1} \partial_{\xi^a}) A_a = -1. \label{eq:A_a_pde}
\end{align}
From equations \eqref{eq:deriv_a_xi_nplus1_xi_n} and \eqref{eq:A_a_pde} we can derive the gradient $A_a$
\begin{align} \begin{split}
	\partial_{\xi^a} A_a &= -\frac{1}{2\xi^{n+1}}(1 + \xi^a A_a) \\
	\partial_{\xi^{n+1}} A_a &= -\frac{1}{2\xi^{n+1}}(A_a + \xi^a \partial_{\xi^a}A_a) \\
	\partial_{\xi^b} A_a &= 0 \text{, when $b=1,\dotsc,n$ and $b\neq a$.} \end{split} \label{eq:pfaffian_of_A_a}
\end{align}

Let $L_a(\vec\xi) = \log A_a(\vec\xi)$ for all $a=1,2,\dotsc,n$. Since $\frac{\partial L_a}{\partial \vec{\xi}} = \frac{1}{A_a} \frac{\partial A_a}{\partial\vec{\xi}}$ we can derive a Pfaffian system for $L_a$,
\begin{align}
	\begin{split}
	\partial_{\xi^a} L_a &= 
	-\frac{1}{2\xi^{n+1}}\left(\frac{1}{e^{L_a}} + \xi^a\right) \\
	\partial_{\xi^{n+1}} L_a &= 
	-\frac{1}{2\xi^{n+1}}(1 + \xi^a \partial_{\xi^a}L_a) \\
	\partial_{\xi^b} L_a &= 0 \text{, when $b=1,\dotsc,n$ and $b\neq a$.},
	\end{split} \label{eq:pfaffian_of_L_a_xi}
\end{align}
and hence we can obtain the gradient of the potential function $\psi^*(\vec\xi) = \sum_{a=1}^n L_a(\vec\xi)$
\begin{align}
\begin{split}
	\partial_{\xi^a}\psi^* &= \partial_{\xi^a} L_a \\
	\partial_{\xi^{n+1}}\psi^* &= \sum_{a=1}^n \partial_{\xi^{n+1}} L_a
\end{split} \label{eq:grad_psi_star}
\end{align}

In addition to the gradient of $\psi^*(\vec\xi)$, we will also need its Hessian, i.e. the matrix of second derivatives, once again as a function of $L_a(\vec\xi)$.
\begin{theorem}
	For any $m \geq 2$ and $a \in [1,n]$, the function $A_a(\vec\xi) = \int_0^\infty \exp(\xi^a y + \xi^{n+1}y^2) \dd y$ satisfies the partial differential equation
	\begin{align*}
		\partial_{\xi^a}^m A_a = -\frac{1}{2\xi^{n+1}} ((m-1)\partial^{m-2}_{\xi^a} A_a + \xi^a \partial^{m-1}_{\xi^a} A_a)
	\end{align*} \label{prop:induction_of_A}
\end{theorem}
\begin{proof}
	The base case $n=2$ is clear from equation \eqref{eq:pfaffian_of_A_a}. Assume $\partial_{\xi^a}^{m-1} A_a = -\frac{1}{2\xi^{n+1}} ((m-2)\partial^{m-3}_{\xi^a} A_a + \xi^a \partial^{m-2}_{\xi^a} A_a)$. Differentiating by $\xi^a$ yields $\partial_{\xi^a}^{m} A_a = -\frac{1}{2\xi^{n+1}} ((m-2)\partial^{m-2}_{\xi^a} A_a + \partial^{m-2}_{\xi^a} A_a + \xi^a \partial^{n-1}_{\xi^a} A_a) = -\frac{1}{2\xi^{n+1}} ((m-1)\partial^{m-2}_{\xi^a} A_a + \xi^a \partial^{m-1}_{\xi^a} A_a)$. \qed
\end{proof}
Now clearly for $a,b=1,\dotsc,n$ and $a \neq b$, we have $\partial_{\xi^a}\partial_{\xi^b} A_a=0$. By \eqref{eq:deriv_a_xi_nplus1_xi_n}, the second derivative of $A_a$ by $\xi^a$ is equal to the derivative by $\xi^{n+1}$. Similarly, $\partial_{\xi^a}\partial_{\xi^{n+1}} A_a = \partial_{\xi^a}^3 A_a$ and $\partial^2_{\xi^{n+1}} A_a = \partial_{\xi^a}^4 A_a$.

Using these we derive the Hessian of $\psi^*$. Again, let $a,b \in {1,\dotsc,n}$ and $a\neq b$. Then
\begin{align*}
	\partial_{\xi^a}\partial_{\xi^b} \psi^* &= 0 \\
	\partial_{\xi^a}^2 \psi^* &= \frac{\partial_{\xi^a}^2 A_a}{A_a} - \left( \frac{\partial_{\xi^a} A_a}{A_a} \right)^2 \\
	\partial_{\xi^a}\partial_{\xi^{n+1}} \psi^* &= \frac{\partial_{\xi^a}^3A_a}{A_a} - \frac{\partial_{\xi^a}^2 A_a}{A_a} \frac{\partial_{\xi^a} A_a}{A_a} \\
	\partial_{\xi^{n+1}}^2 \psi^* &= \sum_{a=1}^n \left[ \frac{\partial_{\xi^a}^4 A_a}{A_a} - \left( \frac{\partial_{\xi^a}^2 A_a}{A_a} \right) \right]
\end{align*}
The Hessian of $\psi^*$ is indeed a function of $\vec\xi$ and $\vec L(\vec\xi) = (L_1(\vec\xi), L_2(\vec\xi), \dotsc, L_n(\vec\xi))^T$, since $\frac{\partial_{\xi^a}A_a}{A_a} = \partial_{\xi^a} L_a$ is a function of $\vec\xi$ and $L_a(\vec\xi)$ by equation \eqref{eq:pfaffian_of_L_a_xi}, and $\frac{\partial_{\xi^a}^m}{A_a}$ is a function of $\frac{\partial_{\xi^a}^{m'}A_a}{A_a}$ for $m\geq 2$, $m' < m$ by Theorem \ref{prop:induction_of_A}, so we can use the holonomic update (see Subsection \ref{sub:holonomic_update_of_the_vector_l_}) to update the vector $\vec L(\vec\xi)$ as $\vec\xi$ changes.


\subsection{Maximum likelihood estimation} 
\label{sub:maximum_likelihood_estimation_trunc_norm}
Next we will discuss details regarding maximum likelihood estimation of the model in equation \eqref{eq:pdf_trunc_norm_theta}. The log-likelihood is easily obtained from equation \eqref{eq:pdf_trunc_norm_theta}
\begin{align}
	\ell(\vec \theta \mid \vec y) = \vec Y^T \mtrx{X}_B \vec\theta - \psi(\vec\theta)
\end{align}
We will use the Holonomic Gradient Method to find the maximum likelihood estimate. Since the Hessian matrix of $\ell(\vec\theta \mid \vec y)$ is easily obtained, we will use the Newton-Raphson method.
Since $\psi(\vec\theta) = \psi^*(\mtrx{X}_B\vec\theta)$, we can use matrix calculus to obtain the Hessian and gradient of the log-likelihood function. Indeed, since the gradient is $\frac{\partial \psi}{\partial \vec\theta} = \frac{\partial \psi^*}{\partial \vec\xi} \cdot \mtrx X_B$ and the Hessian is $\frac{\partial^2 \psi}{\partial \vec \theta \partial \vec{\theta^T}} = \mtrx{X}_B^T \cdot \frac{\partial^2 \psi^*}{\partial \vec \xi \partial \vec{\xi^T}} \cdot \mtrx{X}_B$, we get the gradient and Hessian of the log-likelihood function as follows
\begin{align*}
	(\vec\nabla\ell)^T &= \frac{\partial \ell}{\partial \vec\theta} = \vec Y^T \mtrx{X}_B - \frac{\partial \psi^*}{\partial \vec\xi}\cdot \mtrx{X}_B \\
	H_\ell &= \frac{\partial^2 \ell}{\partial \vec \theta \partial \vec \theta^T} = - \mtrx{X}_B^T \cdot \frac{\partial^2 \psi^*}{\partial \vec \xi \partial \vec{\xi^T}} \cdot \mtrx{X}_B.
\end{align*}

There are some numerical issues to consider when using the method outlined above for maximum likelihood estimation. Let $\vec\theta^{(k)}$ be approximation of the maximum likelihood estimate at the $k$th iteration of the Newton-Raphson method. The next estimate is expressed as $\vec\theta^{(k+1)} = \vec\theta^{(k)} + \vec{\Delta}$, and the difference $\vec{\Delta}$ is obtained by solving the linear system
\begin{align*}
	\mtrx{H}_\ell(\vec\theta^{(k)}) \vec{\Delta} = -\vec\nabla\ell(\vec\theta^{(k)}).
\end{align*}
However, there are times where the Newton-Raphson method is ``too violent'', and yields a $\vec{\Delta}$ of large magnitude, meaning that $\vec\theta^{(k)}$ and $\vec\theta^{(k+1)}$ are relatively far apart. This in turn increases the error in the holonomic update. Furthermore, there are cases where the Newton-Raphson method yields an iterate which does not belong to the model, i.e. when $\theta^{(k+1)}_{d+1} \geq 0$. In our implementation we solve the problem by introducing a small step $\gamma$ when the Newton-Raphson method yields an estimate that is either too far from the previous estimate, or an estimate not belonging to the model.

\subsection{Coordinate conversions} 
\label{sub:coordinate_conversions}
As described in Section \ref{sec:bisector_regression}, we have two sets of e-affine coordinates, $\vec\xi$ and $\vec\theta$, and m-affine coordinates, $\vec\mu$ and $\vec\eta$, along with their potential functions, respectively $\psi^*(\vec\xi)$, $\psi(\vec\theta)$, $\phi^*(\vec\mu)$, and $\phi(\vec\eta)$. The two sets of coordinates are related with
\begin{equation}
\begin{aligned}
	\vec\xi &= \mtrx{X}_B\vec\theta & ~~~ && \vec\eta &= \mtrx{X}_B^T \vec\mu \\
	\psi(\vec\theta) &= \psi^*(\mtrx{X}_B\vec\theta) & ~~~ && \phi(\mtrx{X}_B^T\vec\mu) &= \phi^*(\vec\mu)
\end{aligned} \label{eq:theta2xi_mu2eta}
\end{equation}

Let $P$ be a point on the manifold \eqref{eq:pdf_trunc_norm_theta}, and assume the vector $\vec L(P)$ (the length $n$ vector of the logarithm of normalizing constants of each observation) is known. Given the $\vec \xi$ coordinates of $P$, we can recover its $\vec\mu$ coordinates from equations \eqref{eq:pfaffian_of_L_a_xi} and \eqref{eq:grad_psi_star} since $\mu_i = \frac{\partial \psi^*}{\partial \xi^i}$. Hence
\begin{align}
\begin{split}
	\mu_a &= -\frac{1}{2\xi^{n+1}}\left(\frac{1}{e^{L_a}} + \xi^a\right) \\
	\mu_{n+1} &= -\frac{1}{2\xi^{n+1}} \sum_{a=1}^n (1+\xi^a\mu_a).
\end{split} \label{eq:xi2mu}
\end{align}

We can also invert \eqref{eq:xi2mu} to get the coordinate conversion from $\vec\mu$ to $\vec\xi$
\begin{align}
\begin{split}
	\xi^{n+1} &= -\frac{1}{2(\mu_{n+1} - \sum_{a=1}^n \mu_a^2)}\sum_{a=1}^n \left( 1 - \frac{\mu_a}{e^{L_a}} \right) \\
	\xi^a &= -2\xi^{n+1}\mu_a - \frac{1}{e^{L_a}}
\end{split} \label{eq:mu2xi}
\end{align}

The conversion $\vec\theta$ to $\vec\eta$ is also simple, since we can just compose the transformations in \eqref{eq:theta2xi_mu2eta} and \eqref{eq:xi2mu}, i.e. $\vec\eta(\vec\theta) = \mtrx{X}_B^T \vec\mu(\mtrx{X}_B\vec\theta)$.

Next we will tackle mixed coordinate conversions. As in Subsection \ref{sub:mixed_coordinate_conversion}, let $J \subseteq \{0,1,2,\dotsc, d+r\}$, $\bar{J} = \{0,1,2,\dotsc, d+r\} \setminus J$ and let $P = (\vec\eta_J, \vec\theta^{\bar J})$ denote a mixed coordinate. Additionally, assume that the value of the vector $\vec L$ is known at point $P$.

Newton's method applied to the function F in \eqref{eq:function_mixed2theta_F} will output $\vec\eta_{\bar J}$ and $\vec\theta^J$ at the same time, thus allowing us to recover the full $\vec\theta$ and $\vec\eta$ simultaneously. With the truncated normal distribution, using Newton's method to convert mixed coordinates converges very quickly given a suitable initial guess. Fortunately, there are a few convenient initial guesses that work well. Mixed coordinate conversion is needed in three different situations in the algorithm described in Subsection \ref{sub:helars_algo}:
\begin{enumerate}
 	\item m-projections (steps 2, 4). Use the point before the projection as an initial guess. \\
 	\item updating $\vec L$ (steps 2, 4, 5). Use the point before the update as the initial guess. \\
 	\item the ``wrap-up step'' (step 5). Use the estimate $\hat{\vec\theta}_{(k)}$ of the current iteration as the initial guess.
 \end{enumerate}

 We note again that there are cases where Newton's method outputs a point $(\vec \theta^J,\vec \eta_{\bar J})^{(k+1)} = (\vec \theta^J,\vec \eta_{\bar J})^{(k)} + \vec \Delta^{(k)}$ that does not belong to the model.\footnote{In Newton's method, $\vec\Delta^{(k)} = (\vec\Jac(F))^{-1} F$, where $F$ is the same as in \eqref{eq:function_mixed2theta_F}, and both $F$ and $\vec\Jac(F)$ are evaluated at $(\theta^J, \eta_{\bar J})^{(k)}$} In our implementation, we simply iteratively half the step $\vec \Delta^{(k)}$ until the resulting point $(\vec \theta^J, \vec\eta_{\bar J})^{(k+1)}$ is satisfactory. Such a scaling of the Newton step is required if $d+1 \in J$ and the element $(\theta^{d+1})^{(k+1)}$ in $(\vec \theta^J, \vec\eta_{\bar J})^{(k+1)}$ becomes positive. More precisely, in this case the next iterate becomes
 \begin{align*}
 	(\vec \theta^J,\vec \eta_{\bar J})^{(k+1)} = (\vec \theta^J,\vec \eta_{\bar J})^{(k)} + \left(\frac{1}{2}\right)^\alpha \vec \Delta^{(k)},
 \end{align*}
 where $\alpha = \Biggl\lceil -\frac{\log \left(-\frac{(\theta^{d+1})^{(k)}}{(\Delta^{d+1})^{(k)}}\right)}{\log 2} \Biggr\rceil$.


\subsection{Computational details} 
\label{sub:miscellaneous_implementation_details}
In order to not end up with nearly singular matrices in the algorithm, we will sometimes have to rescale both the design matrix and the response vector. We center and rescale each covariate such that the mean becomes $0$ and the standard deviation becomes $1$. In other words, if $\vec x^i$ is the $i$th column of the design matrix $\mtrx X$, the scaling maps
\begin{align*}
	x^i_j \mapsto \frac{x^i_j - \overline{\vec x}^i}{\sigma_i},
\end{align*}
where $\overline{\vec x}^i = \frac{1}{n}\sum_{j=1}^n x^i_j$ is the mean, and $\sigma^i = \sqrt{\sum_{j=1}^n (x^i_j - \overline{\vec x}^i)^2 / (n-1)}$. Note that as in \cite{hirose2010extension}, scaling and centering and scaling the design matrix will not affect the result of the algorithm. In addition, we will scale the response vector $\vec y$ such that the sample standard deviation equals 1
\begin{align*}
	y_i \mapsto \frac{y_i}{\sigma_y} = \frac{y_i}{\sqrt{\sum_{i=1}^n (y_i - \overline{\vec y})^2 / (n-1)}}.
\end{align*}
These scaling operations allow us to keep the orders of magnitude of the elements in the $\vec\xi$ and $\vec\mu$ coordinates roughly equal, which in turn make the orders of magnitude of the elements in the $\vec\theta$ and $\vec\eta$ coordinates roughly similar. This is needed when doing actual computations, since otherwise many operations involving mixed coordinates (for example the matrix $\mtrx \Jac(F)$ in Proposition \ref{prop:mixed_coord_conv}) will end up nearly singular, with certain columns several orders of magnitude larger than others.


\subsection{Results} 
\label{sub:results}
\begin{figure*}
	\centering
	\includegraphics[width=0.95\textwidth]{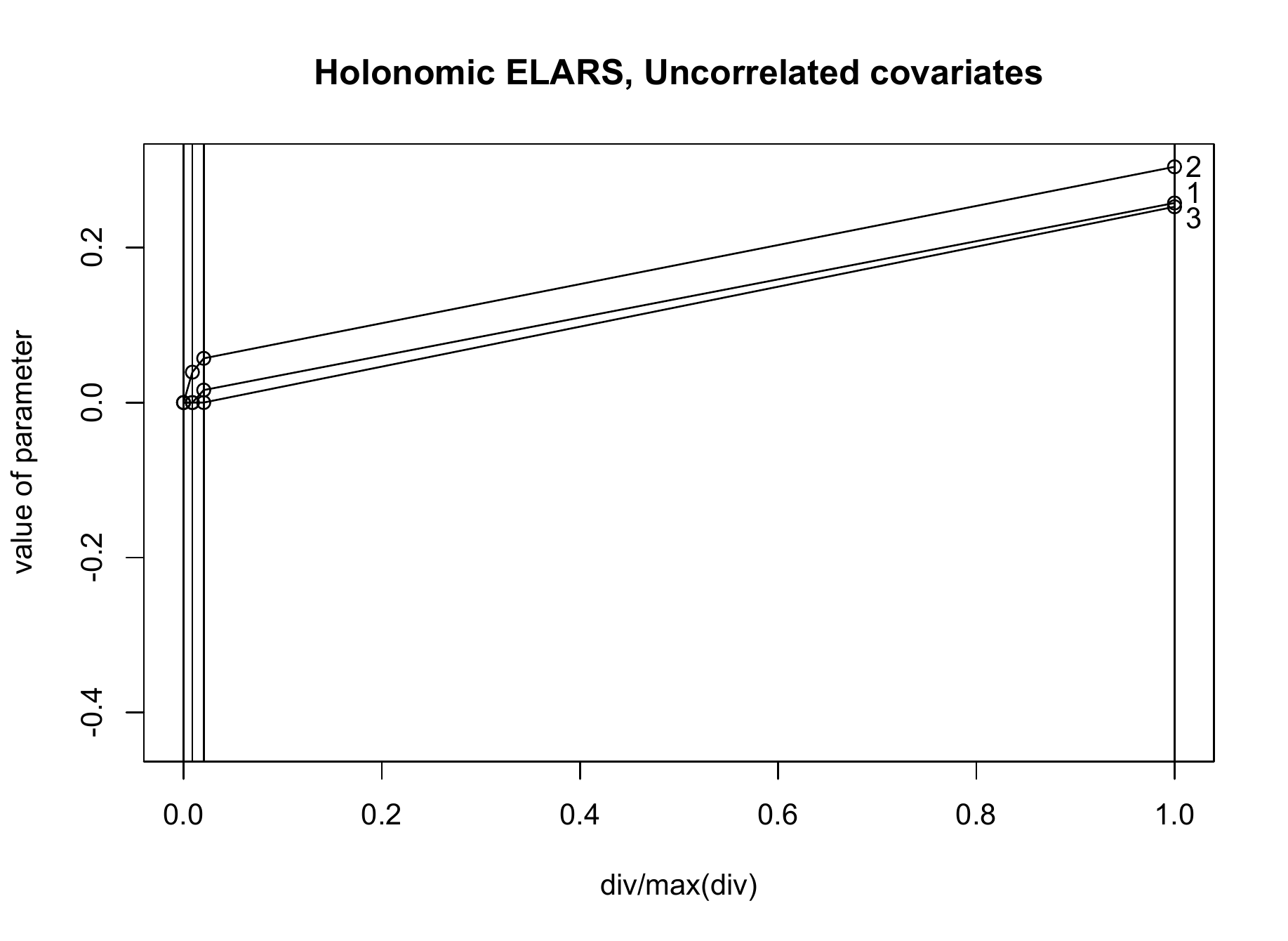}
	\caption{Simulation of $1000$ observations and $3$ uncorrelated covariates.}
	\label{fig:sim_uncor_helars}
\end{figure*}

First, we use a simulated dataset to test the algorithm. We will use $d=3$ covariates $X_1, X_2, X_3$, and $n=1000$ observations. As a first test, we will simulate three uncorrelated covariates. For each observation, each covariate is independently sampled from a uniform distribution between $[0,1]$, and the response is sampled from a truncated normal distribution with mean parameter $X_1 + X_2 + X_3$, and variance $\sigma^2 = 1$. The result of the HELARS algorithm applied to the simulated data is depicted in Figure \ref{fig:sim_uncor_helars}. The algorithm starts on the right, where the value of each parameter is equal to the maximum likelihood estimate of the full model. At each iteration, we compute the divergence of the current parameters compared to the empty model, and we plot the value of each parameter.

The result is as expected: the algorithm sees each covariate as roughly equally important, since they go to zero very close to each other and their value decreases at roughly the same rate. The order in which the covariates go to zero is fully determined by the value of the MLE estimator in the full model. For example since $X_3$ has the smallest coefficient in the full model and it is uncorrelated with the other covariates, it is deemed the least important.

\begin{figure*}
	\centering
	\includegraphics[width=0.95\textwidth]{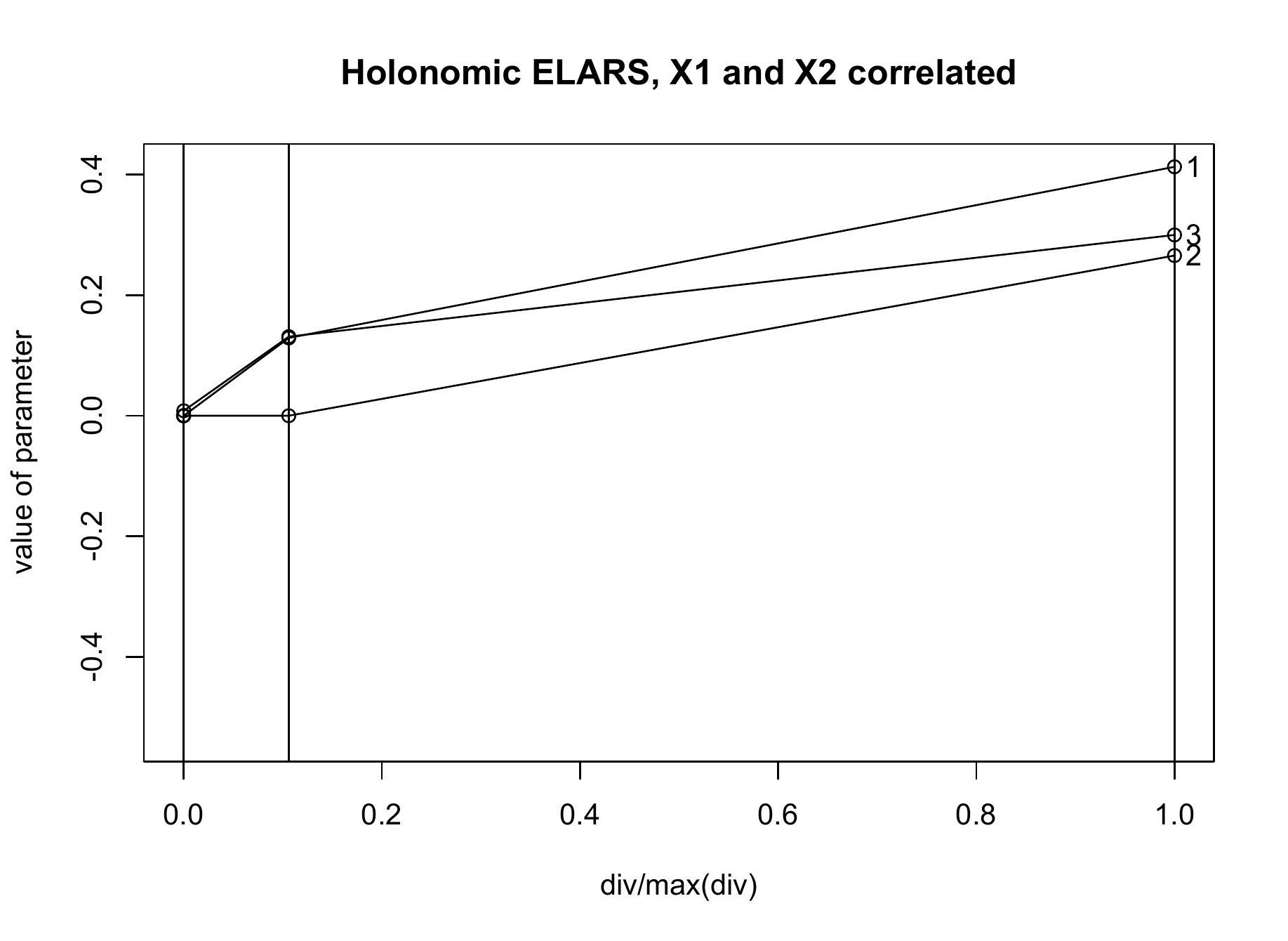}
	\caption{Simulation of $1000$ observations, with covariates $X_1$ and $X_2$ correlated, and $X_3$ independent of the rest.}
	\label{fig:sim_cor_helars}
\end{figure*}

Next, we will introduce correlation between $X_1$ and $X_2$, and leave $X_3$ uncorrelated. The covariates $X_1$ and $X_3$ will once again be sampled from a uniform distribution between $[0,1]$, but $X_2 = X_1 + \varepsilon$, where $\varepsilon \sim N(0, 1/4)$. Again, the response will be sampled from a truncated normal distribution, with mean parameter $X_1 + X_2 + X_3$ and variance parameter $1$. The path of the covariates is in Figure \ref{fig:sim_cor_helars}. We see that $X_2$, one of the two correlated covariates, goes relatively quickly to zero relative to the others, whereas $X_1$ and $X_3$ are deemed to be equally important. One possible interpretation is that $X_2$ is redundant since $X_1$ already carries the same information, so it is quickly eliminated. Once $X_2$ is eliminated, the information of both covariates $X_1$ and $X_3$ is needed, since they are independent. This is also visible when looking at the sum of squared errors (SSE) of each possible subset of covariates in Table \ref{tbl:sim_cor_sse}. Since we know that $X_1$ and $X_2$ are heavily correlated, one of them is redundant and should be removed first. We see that $X_2$ should be removed first, since $\{X_1,X_3\}$ has less error than $\{X_2,X_3\}$. The difference of SSE in the model $\{X_1\}$ and $\{X_3\}$ is due to the fact that the effect of $X_1$ is essentially seen as doubled in the response: recall that the response $Y \approx X_1 + X_2 + X_3$, and since there is a strong positive correlation between $X_1$ and $X_2$, we have $Y \approx X_1 + X_1 + X_3$.

\begin{table}
\caption{Sum of square errors (SSE) using models consisting of every possible subset of covariates. We use a simulation of $1000$ observations, with covariates $X_1$ and $X_2$ correlated, and $X_3$ independent of the rest.}
\label{tbl:sim_cor_sse}
\begin{tabular}{lll}
\hline\noalign{\smallskip}
Subset & $\mathrm{SSE}$ &  $\mathrm{SSE}/\mathrm{SSE_\emptyset}$ \\
\noalign{\smallskip}\hline\noalign{\smallskip}
$\{X_1, X_2, X_3  \}$ &  $725$   & $0.73$ \\
$\{X_1, X_2       \}$ &  $776$   & $0.78$ \\
$\{X_1, X_3       \}$ &  $742$   & $0.74$ \\
$\{X_2, X_3       \}$ &  $770$   & $0.77$ \\
$\{X_1            \}$ &  $792$   & $0.79$ \\
$\{X_2            \}$ &  $824$   & $0.82$ \\
$\{X_3            \}$ &  $948$   & $0.95$ \\
$  \emptyset        $ &  $999$   & $1.00$ \\
\noalign{\smallskip}\hline
\end{tabular}
\end{table}

\begin{figure*}
	\centering
	\includegraphics[width=0.95\textwidth]{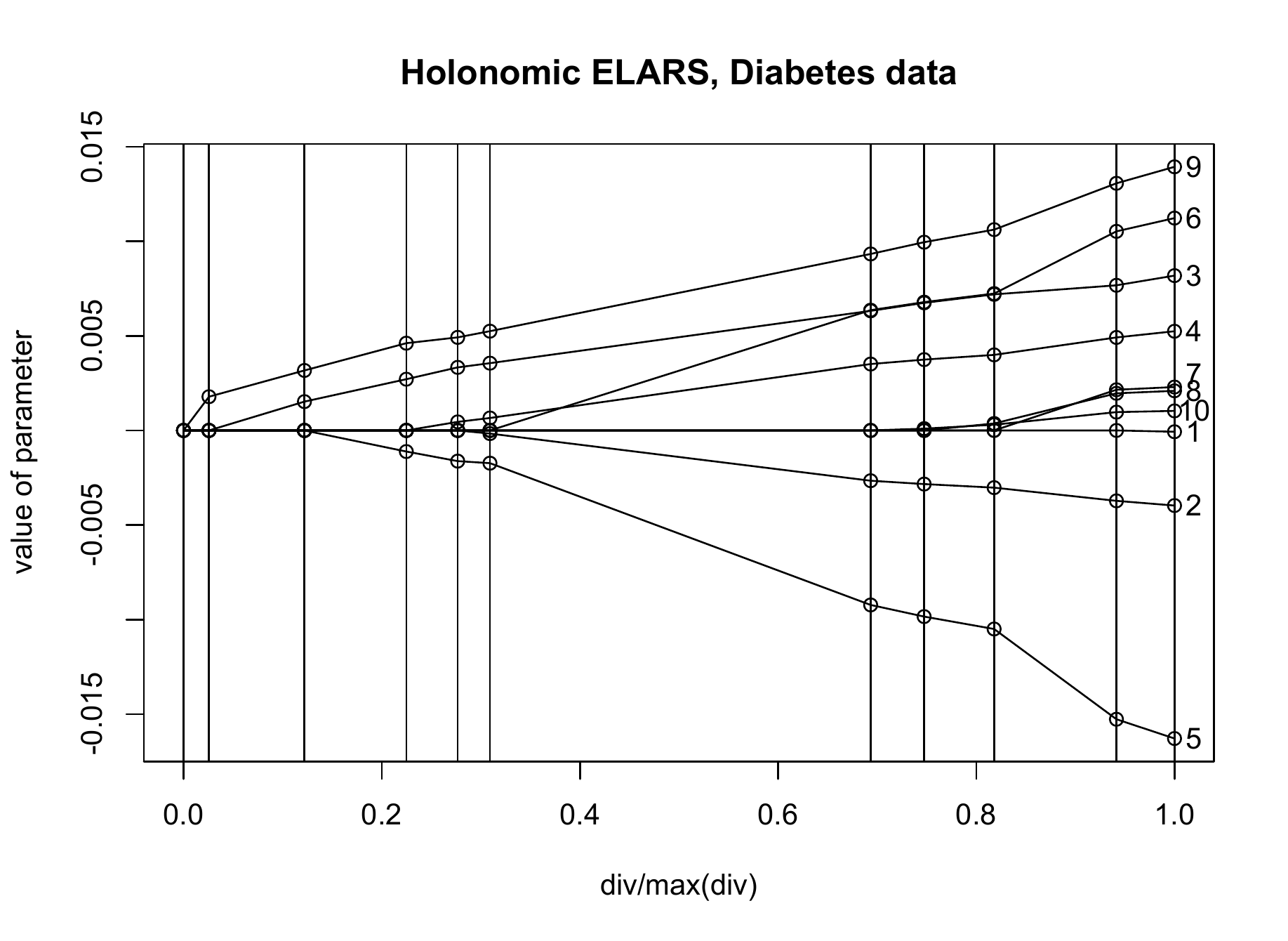}
	\caption{Result of the Holonomic Extended LARS algorithm with the truncated normal distribution on the diabetes data.}
	\label{fig:diabetes_trunc_norm}
\end{figure*}

\begin{figure*}
	\centering
	\includegraphics[width=0.95\textwidth]{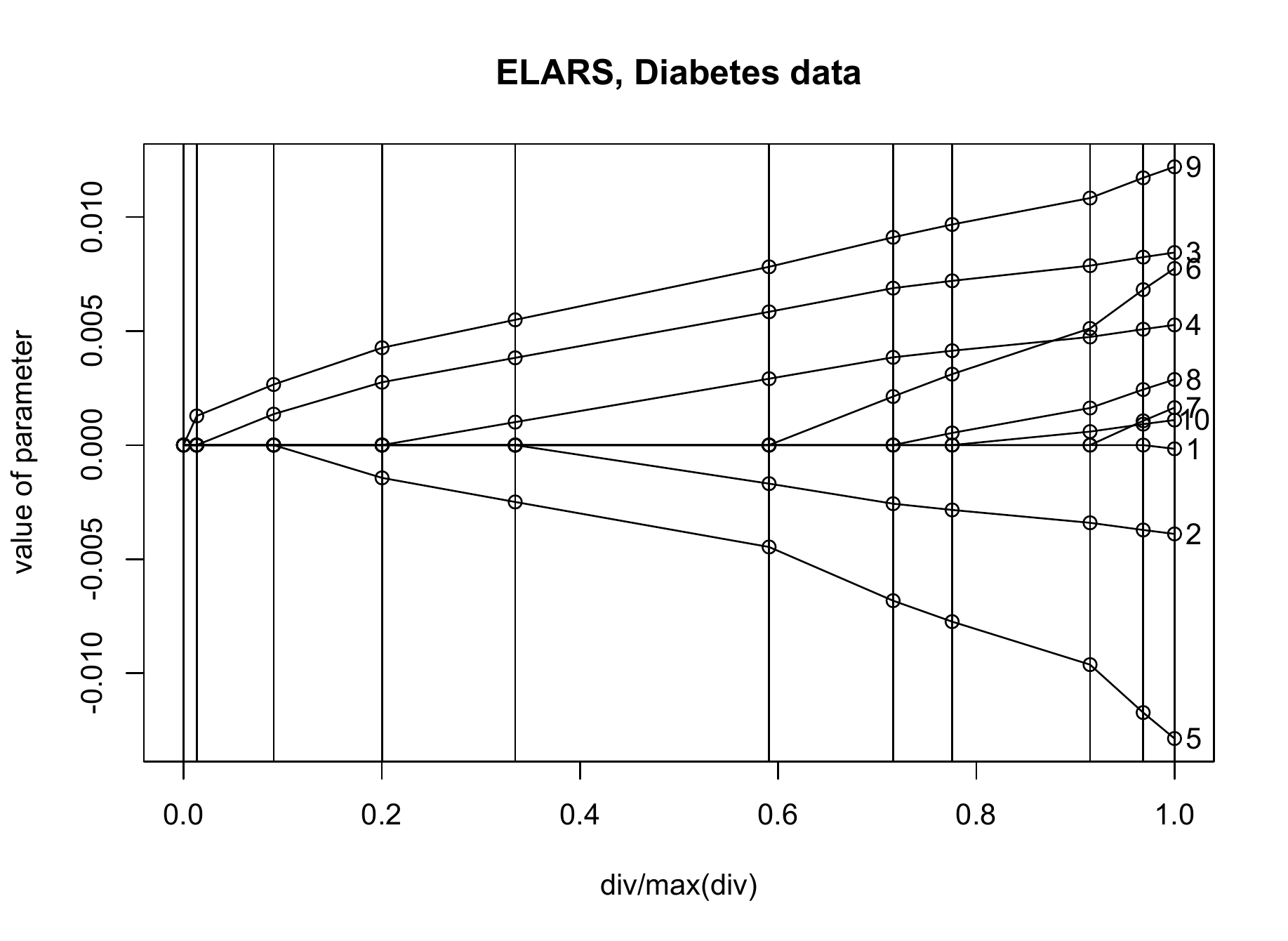}
	\caption{Result of the Extended LARS algorithm with the normal distribution on the diabetes data.}
	\label{fig:diabetes_norm}
\end{figure*}

Next, we used the Diabetes dataset used in the original LARS paper \cite{LARS} and the extended LARS paper \cite{hirose2010extension}. Assuming the truncated normal distribution as the underlying distribution of each observation, the values of $\hat{\vec\theta}$ obtained from the holonomic extended LARS algorithm are plotted in Figure \ref{fig:diabetes_trunc_norm}. The algorithm ordered the covariates in the following order, from least to most important: $\theta_1, \theta_7, \theta_8, \theta_{10}, \theta_6, \theta_2, \theta_4, \theta_5, \theta_3, \theta_9$. We can compare the output of the HELARS algorithm to the output of the ELARS algorithm, depicted in Figure \ref{fig:diabetes_norm}. In the ELARS algorithm we assume that the underlying distribution is the normal distribution, which is why the output looks slightly different. The ELARS algorithm ordered the covariates in the following order: $\theta_1, \theta_7, \theta_{10}, \theta_8, \theta_6, \theta_2, \theta_4, \theta_5, \theta_3, \theta_9$. While the path is different to the truncated normal case, the ordering of variables is almost exactly the same, with the exception of $\theta_8$ and $\theta_{10}$ being flipped.

%% file: Sections/discussion.tex
\section{Discussion} 
\label{sec:discussion}
In this manuscript, we presented the holonomic extended LARS algorithm, and successfully implemented in in \texttt{R}. the dually flat structure is still useful even when the
potential function is not easy to compute. The HELARS implementation is slower than the ELARS implementation due to the overhead caused by keeping track of $\vec L$ and constantly updating it using the holonomic gradient method. The benefits of using holonomicity are most visible when the potential function does not have a closed form expression. Then we can either find a Pfaffian system for the potential function by hand, as we did in our truncated normal distribution example, or use the theory of $D$-modules to construct the Pfaffian system from a holonomic ideal annihilating the potential function. Since in exponential families the potential function is the integral of an exponential function, finding the annihilating ideal is relatively easy in many cases. We can then use the integration algorithm \cite{oaku_algorithms} to get the annihilating ideal of the integral.


Finding the function $G_\psi$ in Equation \eqref{eq:adding_holonicity} satisfying the necessary conditions can also be problematic. At the moment, we have to find it from scratch for every distribution considered. Because finding an elementary enough $G_\psi$ is a very non-trivial task, an algorithm that could automatically output such a function would improve the usability of the HELARS algorithm. Also since the algorithm can only handle a certain class of generalized linear models using the canonical link function, a natural next step would be to extend it to an arbitrary generalized linear model. 